\documentclass[a4paper,11pt]{article}
\usepackage{a4wide}
\usepackage{authblk}
\usepackage{graphicx}
\usepackage{amsmath,amsthm,amssymb}
\usepackage{mathtools}  
\usepackage{listings}

\usepackage[utf8]{inputenc}
\usepackage[swedish,english]{babel}
\usepackage{sidecap, caption}
\usepackage{threeparttable}
\usepackage{subcaption}
\usepackage{algorithm}
\usepackage{algpseudocode}

\usepackage{setspace}

\def\submit{0}
\if\submit1
    \usepackage{lineno}
    \linenumbers
\fi

\if\submit1
\usepackage[backend=biber, style=vancouver, sorting=nyt, sortcites,
firstinits=true]{biblatex}
\else
\usepackage[backend=biber, style=numeric, sorting=nyt, sortcites,
firstinits=true]{biblatex}
\fi

\usepackage[pdftitle={DLCM framework}, pdfauthor={Erik Blom, Stefan
  Engblom}, pdffitwindow=true, breaklinks=true, colorlinks=true,
urlcolor=blue, linkcolor=red, citecolor=red,
anchorcolor=red]{hyperref}

\usepackage{csquotes}
\usepackage{orcidlink}

\numberwithin{equation}{section}
\numberwithin{table}{section}
\numberwithin{figure}{section}

\theoremstyle{definition}

\theoremstyle{plain}
\newtheorem{theorem}{Theorem}[section]
\newtheorem{proposition}[theorem]{Proposition}

\newcommand{\figref}[1]{Fig.~\ref{fig:#1}}
\newcommand{\tabref}[1]{Table~\ref{tab:#1}}
\renewcommand{\algref}[1]{Algorithm~\ref{alg:#1}}

\newcommand{\Nvox}{N_{\mathrm{vox}}}
\newcommand{\Ntypes}{N_{\mathrm{types}}}
\newcommand{\Ncells}{N_{\mathrm{cells}}}

\newcommand{\Omegacomp}{\Omega_{\mathrm{comp}}}
\newcommand{\Omegapop}{\Omega_{\mathrm{pop}}}
\newcommand{\cell}{c_{\mathrm{ep}}}

\newcommand{\Nstoich}{\mathbb{N}}
\newcommand{\Sstoich}{\mathbb{S}}

\DeclareMathOperator{\Expect}{\mathbb{E}}
\newcommand{\Ordo}{\mathcal{O}}
\newcommand{\Ccomplexity}{C_{\mathrm{compl}}}

\begin{document}

\selectlanguage{english}

\title{DLCM: a versatile multi-level solver for heterogeneous multicellular systems}

\if\submit1
\author[1]{Erik Blom\orcidlink{0009-0005-8141-6802}}

\author[1,2*]{Stefan Engblom\orcidlink{0000-0002-3614-1732}}

\affil[1]{{\footnotesize Division of Scientific Computing, Department
    of Information Technology, Uppsala University, Uppsala,
    Sweden.}}

\affil[2]{{\footnotesize Science for Life Laboratory, Department of
    Information Technology, Uppsala University}}


\affil[*]{{\footnotesize Corresponding author. E-mail:
    \href{mailto:stefane@it.uu.se}{stefane@it.uu.se} (SE)}}
\else
\author[1]{Erik Blom\orcidlink{0009-0005-8141-6802}}

\author[1,2*]{Stefan Engblom\orcidlink{0000-0002-3614-1732}}

\affil[1]{{\footnotesize Division of Scientific Computing, Department
    of Information Technology, Uppsala University, Uppsala,
    Sweden. E-mail: \href{mailto:erik.blom@it.uu.se}{erik.blom},
    \href{mailto:stefane@it.uu.se}{stefane@it.uu.se}.}}

\affil[2]{{\footnotesize Science for Life Laboratory, Department of
    Information Technology, Uppsala University}}

\affil[*]{{\footnotesize Corresponding author.}}
\fi


\if\submit1
\date{}
\else
\date{\today}
\fi


\maketitle

\if\submit1
\newpage
\fi

\begin{abstract}
  Computational modeling of multicellular systems may aid in
  untangling cellular dynamics and emergent properties of biological
  cell populations. A key challenge is to balance the level of model
  detail and the computational efficiency, while using physically
  interpretable parameters to facilitate meaningful comparisons with
  biological data.

  For this purpose, we present the DLCM solver (discrete Laplacian
  cell mechanics), a flexible and efficient computational solver for
  spatial and stochastic simulations of populations of cells,
  developed from first principle to support mechanistic
  investigations. The solver has been designed as a module in URDME,
  the unstructured reaction-diffusion master equation open software
  framework, to allow for the integration of intra-cellular models
  with extra-cellular features handled by DLCM. The solver manages
  discrete cells on a fixed lattice and reaction-transport events in a 
  continuous-time Markov chain. Space-continuous micro-environment 
  quantities such as pressure, nutrients, and chemical substances are 
  supported by the framework, permitting a variety of modeling choices
  concerning chemotaxis, mechanotaxis, nutrient-driven cell growth and
  death, among others.

  An essential and novel feature of the DLCM solver is the coupling of
  cellular pressure to the curvature of the cell populations by
  elliptic projection onto the computational grid. This provides a
  consistent evaluation of population curvature with which we can
  include effects from surface tension between different
  populations---an essential mechanism in, e.g., problems in
  developmental biology.
  
  We demonstrate the flexibility of the modeling framework by
  implementing benchmark problems of cell sorting, cellular signaling,
  tumor growth, and chemotaxis models. We additionally formally
  analyze the computational complexity and show that it is
  theoretically optimal for systems based on pressure-driven cell
  migration. In summary, the solver strikes a balance between
  efficiency and a relatively fine resolution, while simultaneously
  supporting a high level of interpretability.

\bigskip
\noindent
\textbf{Keywords:} Cell population modeling, Darcy's law, Cell
signaling, Stochastic modeling, Reaction-Diffusion system.

\medskip
\noindent
\textbf{AMS subject classification:} \textit{Primary:} 92-04, 92-08,
92-10; \textit{secondary:} 92C15, 65C40, 60J28.

%
%

\if\submit0
\medskip
\noindent
\textbf{Funding:} This work was partially funded by support from the
Swedish Research Council under project number VR 2019-03471.  \fi

\end{abstract}


\if\submit1
\subsection*{Author Summary}


Computational models of populations of cells can elucidate connections
between the behavior of individual cells and the emergent properties
of the population they comprise. To draw meaningful conclusions from
such simulations, however, comparable biological data must exist that
support both the model parameters and the outcome. Another challenge
is to balance the complexity of the model with the computational
resources required to simulate it.

To tackle these challenges, we have developed a framework for
simulating cell populations that aims to balance computational
efficiency and physical realism, while still being capable of handling
a wide variety of biological phenomena. The framework treats the cells
as agents represented by state vectors on a fixed lattice embedded in
an environment of nutrients and chemicals, where each cell is equipped
with an internal state detailing, e.g., protein levels and
phenotype. The latter are, together with cell movement, birth, and
death, treated as stochastic events, thus incorporating biological
noise into the framework. The framework also manages population-level
surface tension effects, employing a novel technique in the context.

We demonstrate the utility of the framework through a variety of cell
population models that explore cell sorting and signaling, tumor
growth, and cell migration driven by chemical cues.
\fi


\section{Introduction}
\label{sec:introduction}


Computational modeling provides an excellent complement to biology
research by testing hypotheses, uncovering mechanistic and causal
relations, and motivating new experiments, among other uses
\cite{brodland2015computational}. When examining the self-organization
and the emergent behavior of multicellular tissues, there are many
different modeling approaches at the modeler's disposal, ranging from
the continuum dynamics of ordinary and partial differential equations
(ODEs and PDEs, respectively), to agent-based models (ABMs) such as
cellular automata, cellular Potts, center-based, and vertex models.

The diversity of the available tools reflects the complexity of the
biological processes being studied. For example, tumor dynamics are
intricately linked with the tumor's micro-environment---necessitating
models that can capture this interaction
\cite{macklin2016progress}. One of the many challenges in the
computational modeling of multicellular systems lies in ensuring
meaningful comparison and fully understanding the consistency between
the different modeling approaches, assumptions, and subsequent
software implementations \cite{fletcher2022sevenchallenges,
  macklin19keychallenges}. With this paper we aim to contribute to
this quest by proposing a mechanistically founded simulation framework
which is efficient enough to support simulation-based discovery and
also facilitates mathematical model analysis.

ABMs are useful when modeling heterogeneous populations where the
resolution of the individual cell (or other biological agents) is
necessary \cite{pleyer23abm_cellular}. Because of this, many models
and frameworks have been developed and used with this capacity in
mind, e.g., Chaste is capable of handling many of the standard ABM
types \cite{Chaste}, CompuCell3D supports cellular Potts-type models
\cite{CompuCell3D}, and PhysiCell center-based models
\cite{PhysiCell}, see also the comparisons of these and other
frameworks in \cite{pleyer23abm_cellular}.

While platforms such as these provide a means to tackle some of the
challenges to modeling multicellular systems (for examples, see
\cite{fletcher2022sevenchallenges}), each have their own inherent
trade-offs. For example, cellular Potts models offer flexibility in
incorporating new mechanisms, but face ambiguities in behavior, time
scale, and parameter interpretation \cite{anja2012cpm}. Center-based
models, while based on physically interpretable cell-cell forces and
interactions, are computationally expensive and often overly detailed
compared to supporting biological data. Rule-based cellular automata
models, finally, while efficient and facilitating insightful
mathematical treatment, can suffer from poor physical interpretability
and unwanted lattice-based artifacts \cite{osborne2017comparing}.

Motivated by this, we have previously proposed an event-based
framework dubbed discrete Laplacian cell mechanics
\cite{engblom2018scalable}, which as the name suggests evolves cells
relying on discrete Laplacian operators. We discuss here the further
development and design of the \emph{DLCM solver}, a novel contribution
to the URDME open software framework
\cite[\url{www.urdme.org}]{URDMEpaper}. Concretely, we present herein
a general version of the framework formalism from the previous more
specific design and we develop a first fully working implementation
together with some theoretical results covering computational
properties. The solver is designed to balance flexibility, efficiency,
and interpretability in a stochastic spatial event-driven modeling
approach. Based on first principle reaction-transport physics, DLCM
models heterogeneous cell populations in a continuous-time Markov
chain (CTMC). Each cell inhabits a voxel in a computational grid --
which can be structured or unstructured -- and is governed by reaction
and transport dynamics based on local rules and continuum micro
environment quantities, such as oxygen, pressure, and chemical
signals, each modeled as a Laplacian equilibrium.

The balance between various trade-offs in the DLCM framework differs
from known frameworks in various ways. Unlike off-lattice models like
Vertex and Voronoi \cite{meineke2001voronoi}, DLCM does not keep track
of a continuous representation of cell shape or position in fine
detail, allowing for efficient simulations of models whenever the
effects of such details are negligible. The cellular mechanics in the
framework is based on continuous physics facilitating a physical
interpretation as well as mean-field behavior analysis
\cite{blom2024morphological}. This is in contrast to, e.g., cellular
Potts models where certain non-local mechanisms severely complicate
such analysis \cite{anja2012cpm}, or general cellular automata for
which a well-defined continuous physics in the limit is not guaranteed
under given update rules. In line also with other frameworks for
multicellular systems, DLCM allows any time-continuous effects to be
added in a consistent way, using individual cell data for
heterogeneous populations, and continuous steady state quantities for
the micro-environment.

In this paper, we develop the DLCM solver details in
\S\ref{sec:solver}, followed by a suite of test cases using the solver
in \S\ref{sec:results} intended to highlight its expressiveness,
interpretability, and ability to reproduce behaviors in a selection of
benchmark models. Motivated in part by the framework comparisons in
\cite{osborne2017comparing}, we apply the DLCM framework to the cell
sorting benchmark and Hes1 signaling model in addition to models of
tumor growth and a variety of chemotaxis models in two and three
spatial dimensions, demonstrating the solver's capacity to handle a
wide range of problems. The solver simulations reproduce the
pattern-formation in the sorting and signaling experiments and
exemplify physical interpretability using the tumor model. Notably,
the proposed treatment of surface tension is a novel contribution in
the context and offers a mechanical interpretation at the macroscopic
level. While we do not yet utilize the full potential of the
framework's efficiency, in \S\ref{sec:solver_complexity} we do include
a theoretical derivation of the solver's computational complexity
which indicates optimal performance in the setting. Here we also
assess in what sense the solver is consistent with the model
physics. We discuss the potential and limitations of the solver
alongside future developments in \S\ref{sec:discussion}.


\section{Materials and Methods}
\label{sec:solver}

Here we first offer a general overview of the DLCM framework and its
multi-layered structure. A more formal presentation of the class of
models the solver handles is found in \S\ref{sec:framework_formalism}
together with a technical development on how surface tension effects
are effectively managed in \S\ref{sec:curvature_surfacetension}. In
\S\ref{sec:implementation_details}, we outline the details of the
solver implementation, and in \S\ref{sec:solver_complexity} we discuss
in what sense the modeling is consistent with known physics as well as
the resulting solver's computational complexity under quite general
model assumptions.

\subsection{DLCM framework overview}
\label{sec:DLCM_overview}


Similar to certain other frameworks, e.g., PhysiBoss
\cite{letort2019physiboss}, DLCM supports simulation at the following
three scales: the population micro-environment, the cell mechanics and
phenotypic behavior, and the intracellular dynamics. More
specifically, cells in the DLCM framework are understood as discrete
agents represented by a state vector that move, proliferate, die, or
switch phenotype based on event intensities or \emph{propensities} in
a continuous-time Markov chain (CTMC). The cells reside in a
computational grid and are equipped with internal states which can be
either continuous or discrete entities, such as quantities of proteins
or signaling molecules, and that change according to its own set of
internal propensities. Finally, the micro-environment is modeled by
quasi-stationary PDEs, whose fields may interact with the cells.

\begin{figure}[H]
  \centering
  \includegraphics{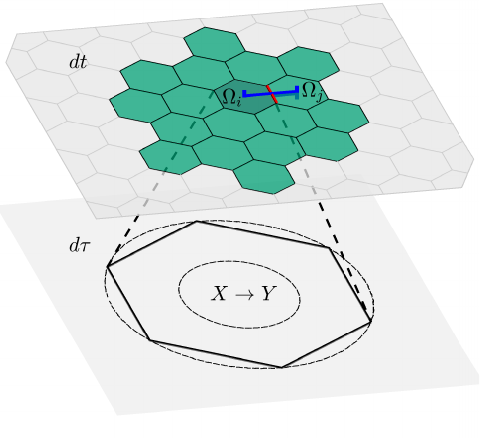}
  \caption{
    \if\submit1
    \doublespacing
    \fi
    \textbf{The two-level structure of the DLCM
      framework}. \textit{Top:} the outer population level where cells
    are represented by their state vector residing in discrete voxels
    on a grid, and whose position and phenotype change by state update
    rules in time-increments of scale $dt$. Green represents occupied
    voxels. The blue line connects the centers of the two voxels,
    $\Omega_i$ and $\Omega_j$, and the red line is their shared edge:
    $d_{ij}$ and $e_{ij}$, respectively, in
    \eqref{eq:rates_migration_explicit} below. \textit{Bottom:} the
    inner cellular level where internal reactions, such as protein
    synthesis or decay (here represented by $X\rightarrow Y$), occur
    in time increments of scale $d\tau$, generally assumed to be
    smaller than the external $dt$.}
  \label{fig:2layer}
\end{figure}

A foundational assumption of the framework is that the internal states
of the cells only affect the population-level events in a weak
sense. A model that breaks this assumption would be one where the
internal cell dynamics -- such as the synthesis of a protein -- vary
on a comparably fast scale \emph{and} strongly feed back into the
population-level events, thereby altering the timescale of the
latter's evolution. We note that this foundational assumption does not
exclude the cell population events to depend on the internal states,
e.g., migration rates which depend on a cell-specific chemotactic
sensitivity, but only that the resulting feedback effect cannot be so
strong that it alters the relative difference in timescales. In
effect, this scale separation assumption allows the population-level
state to be regarded as approximately constant during the inter-event
times of the cells themselves and naturally suggests the multi-level
structure proposed in \cite{engblom2019stochastic}, in which two
distinct computational layers handle the external and internal rates
separately. Both layers update asynchronously through stochastic
simulation algorithm (SSA) routines, first sampling the waiting time
and event of the population layer, then updating each cell's
internal state in the cellular layer during said waiting time
(cf.~\figref{2layer}). Continuous internal states are naturally
updated in continuous time using standard ODE solvers 
during the waiting time. A key feature of this multi-level coupling
is that both layers work in continuous time, preserving the
interpretability of time in the resulting framework.

Events at both levels of description may depend on continuous
micro-environment quantities, e.g., nutrients, oxygen, and cellular
pressure, that are all modeled by stationary Poisson equations and are
solved over the whole population of cells. The rationale behind this
design is that the dynamics of the micro-environment are typically
much faster than that of the cell population and, consequently, can be
modeled as if it reaches stationarity immediately after any cell
population event. The cellular pressure is coupled to the population
geometry via boundary conditions to incorporate the effect of surface
tension.

When regarded as an URDME solver, the DLCM framework thus offers three
new distinct capabilities: \textit{(i)} it introduces agents with
individual information and dynamic phenotypes that characterize
behavior, \textit{(ii)} it adds global quantities which affects the
cellular movements as well as other events, and \textit{(iii)} it
supports an evaluation of population curvature and surface tension
effects.

We also emphasize that the dynamics of the DLCM framework, like URDME,
are fundamentally stochastic due to the underlying Markov chain.
Therefore, the noise both intrinsic and essential to many biological
mechanisms at the cell and population level is intrinsic also to the
framework. It follows that the implicitly defined likelihood could be
relied upon for inference using likelihood-based methods.

\subsection{Framework formalism}
\label{sec:framework_formalism}


%
The computational domain, $\Omegacomp$, is discretized into
$i = 1, 2, ..., \Nvox$ voxels $\Omega_i$ forming the computational
mesh $\Omega_h$. The DLCM solver can formally be used over any grid
for which a consistent discrete Laplace operator can be derived;
however, we have focused on finite element discretizations which
produce flexible discrete operators and support a rigorous
theory. Each voxel may contain a number of cells, $u_i$ from zero up
to $2$ (cf.~\S\ref{sec:implementation_details}); each cell is equipped
with a label $l = 1,...,\Ntypes$ that represents cell phenotype and a
local data structure with individual cell information. The latter is
updated by independent internal reactions per cell on a generally
shorter time scale, cf.~\figref{2layer}~(\textit{bottom}).

The framework physics, however, is formulated from continuous
assumptions and laws. Letting $u = u(x,t)$ represent the cell
distribution as a continuous density of cells, reaction-transport
dynamics govern the cells according to
\begin{equation}
    \label{eq:external_layer_physics}
    \frac{\partial u}{\partial t} + \nabla \cdot I = \Nstoich 
    \omega(u,v,w),
\end{equation}
where $I$ is the cell flux, $v = v(x,t)$ is the micro-environment
(pressure, nutrients, etc.), $\Nstoich$ is the stoichiometric matrix
detailing reactant and product information (for birth, death, and
phenotype switching events), and $\omega$ the vector of respective
event rates. In the continuous setting, we let each internal cell
state be represented as a concentration at each point, $w = w(x,t)$,
with independent reaction rates, $\nu(u,v,w)$, and a stoichiometric
matrix $\Sstoich$,
\begin{equation}
    \label{eq:internal_layer_physics}
    \frac{\partial w}{\partial t} + \nabla \cdot I = \Sstoich\nu(u,v,w),
\end{equation}
using the same flux, $I$, as for the cells in
\eqref{eq:external_layer_physics}, in this way formally expressing
that the states $w$ pertain to the moving cells.

In essence, the external solver layer \figref{2layer}~(\textit{top})
handles \eqref{eq:external_layer_physics} and the internal layer
\figref{2layer}~(\textit{bottom}) handles the reactions in
\eqref{eq:internal_layer_physics}, where respective layer dynamics are
computationally decoupled. The implementation details of the
decoupling, the discretization and sampling of the reactions
represented here by $\omega$ and $\nu$ are described in
\S\ref{sec:implementation_details}. Subsequently, we characterize the
general physics governing the micro-environment quantities $v$ and the
flux $I$.

The framework handles an arbitrary number of field
variables describing the population micro-environment, denoted $v^{(m)} =
v^{(m)}(x,t)$, for $m=0, 1, 2,...,M$. These are modeled by stationary
diffusion equations on the general form
\begin{align}
 \label{eq:laplace_quantities}
 \begin{split}
    -\Delta v^{(m)} &= s^{(m)}(u, v, w), \\
     v^{(m)} &= r^{(m)}(u, v, w),
     \quad \text{at } \partial \Omegacomp,
 \end{split}
\end{align}
including also a possible explicit dependence on time and space which
we omit for brevity, and where $v = [v^{(0)}, v^{(1)}, ...,
v^{(M)}]$. The micro-environment quantities in
\eqref{eq:laplace_quantities} fundamentally affect the reactions rates
$\omega$ and $\nu$, but also the cellular migration as described
next.

Cell migration is based on the physical flux of cell densities in
\eqref{eq:external_layer_physics} and is
proportional to the gradient of some function of $v$ and the cellular
properties. The framework supports such cell migration arising from
several different sources according to the combined flux
\begin{align}
  \label{eq:migration_physics}
  I = \sum_n D^{(n)}(u, v, w) \nabla f^{(n)}(u,v,w),
\end{align}
for $n$ up to an arbitrary number of flux terms, where
$D^{(n)}$ is the flux scaling factor, and the
$f^{(n)}$ are the \textit{migration potentials}, both possibly with
an explicit dependence on time and space as well. The discretization
of the flux and the event-based interpretation of cell migration is
detailed in \S\ref{sec:implementation_details}.

As an example of cell migration, this approach directly supports the
modeling of cell migration by Darcy's law \cite{whitaker1986flow}
\begin{equation}
  \label{eq:darcys_law}
  I = - uD \nabla p,
\end{equation}
where $D$ is the Darcy coefficient, and $p$ is the cellular pressure
modeled continuously according to \eqref{eq:laplace_quantities}. We
specifically reserve the index $m = 0$ for this pressure, i.e.,
$p \equiv v^{(0)}$.  Other forms of gradient-dependent migration, such
as chemotaxis, are readily handled by the solver in a similar manner
(cf.~\S\ref{sec:chemotaxis}), but pressure is specifically included to
handle effects of surface tension, as discussed in the next section.

\subsection{Curvature and surface tension}
\label{sec:curvature_surfacetension}


In this section we assume some familiarity with finite element
methods, following the notation in the monograph
\cite{larson2013finite} closely.

How heterotypic cell populations sort themselves is a key aspect in
the development of tissues and organs \cite{Gomezgalvex2021}. There
are many hypotheses concerning the nature of cells' self-organization
and sorting in developmental biology \cite{Krens2011_cellsorting,
  tsai2022selforgreview, foty2005differential}, some of the most
prominent concerning cell-cell adhesion and surface tension and which
are amenable to scrutiny using computational modeling
\cite{osborne2017comparing}. To facilitate the study of such models,
the DLCM framework is capable of handling curvature and adhesion
effects. Surface tension is included in the evaluation of pressure
through Young-Laplace pressure drops between population interfaces, as
in
\begin{equation}
  \label{eq:young-laplace}
  p(x^+) - p(x^-) = \sigma_{kl} C,
\end{equation}
where $x^\pm$ denotes points at the interface approached from within
population $k$ and $l$, respectively, $\sigma_{kl}$ is the
phenomenological surface tension coefficient between the cells of type
$k$ and $l$, and $C$ is the interface curvature. The condition
\eqref{eq:young-laplace} has been used in other models of cell
populations of, e.g., wound healing \cite{ben2014re-ep} and bacterial
colony growth \cite{giverso2016emerging}.

The curvature $C$ in \eqref{eq:young-laplace} is evaluated from the
continuous interface via
\begin{equation}
    \label{eq:divergence}
  C = \nabla \cdot n,
\end{equation}
where $n$ is the interface normal vector
\begin{equation}
  \label{eq:normalized_normal}
  n = \frac{\nabla u}{\lVert \nabla u \rVert_{l^2}},
\end{equation} 
and where $\|\cdot\|_{l^2}$ denotes the discrete Euclidean norm of the
gradient vector.

We propose here to approximate the continuous curvature $C$ on the
discrete grid in three steps: \textit{(i)} imposing smooth gradients
of $u$, \textit{(ii)} evaluating \eqref{eq:normalized_normal}, and
finally \textit{(iii)} evaluating \eqref{eq:divergence}; all performed
using \textit{elliptic projection} onto the discrete finite element
space $V_h$ associated with the computational mesh
$\Omega_h$. Generally speaking, when projecting a function $f \in V$
in some suitable Sobolev function space $V$, we seek the function
$\hat{\phi} \in V_h \subseteq V$, where $V_h$ is a discrete subspace
of $V$. In our context, $V_h$ is the space spanned by a finite element
method's basis functions over the elements in $\Omega_h$. The
projection is constructed to minimize the error $\hat{\phi}-f$ in an
appropriate norm:
\begin{equation}
  \label{eq:projection_minimizer}
  \lVert \hat{\phi} - f \rVert \leq \lVert v-f
  \rVert ,
  \quad \forall v \in V_h.
\end{equation}
With $\lVert \cdot \rVert$ induced by an inner product,
$(\cdot , \cdot)$, \eqref{eq:projection_minimizer} can be solved in
variational form,
\begin{equation}
  \label{eq:L2_projection}
  (\hat{\phi}-f, v) = 0,
  \quad \forall v \in V_h,
\end{equation}
where we rely on the usual inner product for functions $f,g \in V$,
i.e.,
\begin{align}
  \label{eq:inner_product}
  (f,g) &\equiv \int_{\Omegacomp} f(x) g(x) \, dx.
\end{align}

Unfortunately, \eqref{eq:projection_minimizer} lacks control over the
gradient of the projection, potentially rendering the gradient
estimates in subsequent steps meaningless for our purposes.  Using
elliptic projection instead, we add a regularizing term to the
variational form to obtain
\begin{equation}
  \label{eq:elliptic_projection}
  (\hat{\phi}-f, v) + \cell h^2(\nabla \hat{\phi}, \nabla v) = 0,
  \quad \forall v \in V_h,
\end{equation}
where $\cell$ is a smoothing coefficient and $h$ is some appropriate
representation of mesh size, e.g., the maximum edge length
$h_{\mathrm{max}}$. Then, we have the following result.
\begin{proposition}[\textit{Elliptic projection}]
  \label{prop:elliptic_projection_minimizer}
  The function $\hat{\phi}$ given by \eqref{eq:elliptic_projection} satisfies
  \begin{equation}
    \label{eq:elliptic_minimizer}
    \lVert \hat{\phi} - f \rVert^2 + \cell h^2
    \lVert \nabla \hat{\phi} \rVert^2 \leq \lVert v - f \rVert^2 + \cell h^2
    \lVert \nabla v \rVert^2, \quad \forall v \in V_h,
  \end{equation}
  which reduces to \eqref{eq:projection_minimizer} for $\cell =
  0$. Further, assuming the standard linear basis functions
  $\{\varphi_i\}_{i=1}^{n}$ over the elements in $\Omega_h$, the
  solution coefficient vector $\xi = [\xi_1,\ldots,\xi_n]^T$ to
  \eqref{eq:elliptic_projection}, i.e., by which
  $\hat{\phi}(x) \coloneqq \sum_{i = 1}^n \xi_i \varphi_i(x)$,
  satisfies the linear system
  \begin{equation}
    \label{eq:elliptic_scheme}
    (M + \cell h^2 A)\xi = b,
  \end{equation}
  with mass matrix
  $M_{ij} = \int_{\Omegacomp} \varphi_j \varphi_i \, dx $, stiffness
  matrix
  $A_{ij} = \int_{\Omegacomp} \nabla \varphi_j \cdot \nabla\varphi_i
  \, dx$, and load vector $b_i = \int_{\Omegacomp} f \varphi_i \, dx$,
  for $i,j = 1, 2, ..., n$.
\end{proposition} 
As \eqref{eq:elliptic_minimizer} shows, $\cell$ imposes a penalty on
the gradient of the projected function, facilitating control over the
smoothness of the estimates of the population gradient and its
curvature.  We provide a proof of
Proposition~\ref{prop:elliptic_projection_minimizer} in
\S\ref{apx:proof}.

The algorithm for approximating the curvature $C$ of the population
boundary of $u$ entails three applications of
\eqref{eq:elliptic_projection} using the numerical scheme
\eqref{eq:elliptic_scheme}: \textit{(i)} we obtain a smoother
representation of the cell population, denoted $\hat{u}$, by setting
$f = u$ in \eqref{eq:elliptic_projection} since sharp gradients will
destroy the estimates of the normal and curvature in the next steps;
\textit{(ii)} we set $f = \nabla \hat{u}$ and normalize the result
according to \eqref{eq:normalized_normal} to get $\hat{n}$;
\textit{(iii)} setting $f = \nabla \cdot \hat{n}$, we get the
approximate curvature $\hat{C} \in V_h$ by solving
\eqref{eq:elliptic_projection} a third and final time. To get the
curvature for all cell types for which \eqref{eq:young-laplace} is to
be imposed, we perform the three steps for each such population
separately. For voxels containing mixed cell types, we make use of the
pragmatic convention that the ``first'' cell in the voxel determines the
cell type for the purpose of evaluating the population curvature. This
is also the cell that is next to move according to the first-in,
first-out principle detailed in the next section. To get the curvature
between two populations of phenotype $k$ and $l$ we use the average
$\hat{C} = 0.5 \times (\hat{C}_k + \hat{C}_l)$ before applying
\eqref{eq:young-laplace} lest we introduce a discretization
bias. \algref{elliptic_proj} summarizes the elliptic projection
procedure.

{\centering
  \begin{minipage}{1\linewidth}
    \begin{algorithm}[H]
    \if\submit1
        \doublespacing
    \fi
      \caption{Curvature by elliptic projection.}
      \label{alg:elliptic_proj}
      \begin{algorithmic}
        \For{$1 \le n \le \Ntypes$}
        \State For cell population $u$ of type $n$:
        \State $f \gets u$; find $\hat{\phi}_1$, such that 
        \eqref{eq:elliptic_projection}. \Comment{\textit{(i)} smooth population}
        \State $f \gets \nabla \hat{\phi}_1$; find $\hat{\phi}_2$, such that 
        \eqref{eq:elliptic_projection}. \Comment{\textit{(ii)} get population gradient}
        \State $\bar{\phi}_2 \gets \hat{\phi}_2 / \lVert \hat{\phi}_2
        \rVert_{l^2}$ \Comment{normalize}
        \State $f \gets \nabla \cdot \bar{\phi}_2$; find $\hat{\phi}_3$, such that 
        \eqref{eq:elliptic_projection}. \Comment{\textit{(iii)} get curvature}
        \State $\hat{C} \gets \hat{\phi}_3$
        \State Keep only the curvature values $\hat{C}$ at the current
        population interface. 
        \EndFor
      \end{algorithmic}
    \end{algorithm}
  \end{minipage}
}

\begin{figure}[ht!]
  \centering
  \includegraphics[width=1\linewidth]{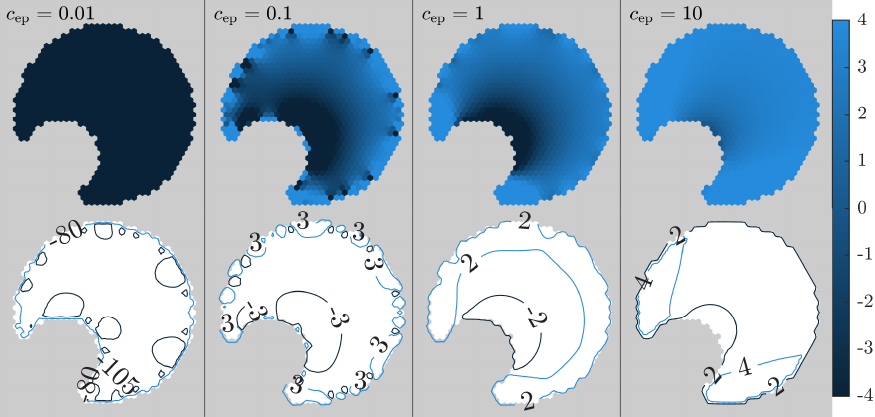}
  \caption{
    \if\submit1
    \doublespacing
    \fi
    \textbf{Impact of elliptic projection penalty parameter on the
      curvature evaluation}. The pressure distribution (\textit{top
      row}, colormap range capped to $[-4,4]$) and corresponding
    isolines (\textit{bottom row}) for a circular population with a
    smaller circle removed from it. The boundary values are equal to
    the curvature, i.e., $\sigma_{10} = 1$ and $p=0$ outside the
    population, cf.~\eqref{eq:young-laplace}, and the curvature is
    estimated from \algref{elliptic_proj}. From left to right,
    $\cell = [0.01, 0.1, 1, 10]$, the second value being used in the
    experiments in \S\ref{sec:results}. The rightmost example shows no
    negative curvature values inside the smaller circle, and the
    leftmost shows negative, irregular values across the entire
    boundary.}
  \label{fig:curvature}
\end{figure}

The coefficient $\cell$ can be adjusted for more or less smoothing of
the curvature, but the recommended range is $\cell \in (0.05,5)$ to
avoid excessive noise or insensitivity to small, but relevant, changes
in the geometry. \figref{curvature} shows the impact of $\cell$ on the
curvature of a test population using \algref{elliptic_proj}. Given two
values outside the recommended range, the curvature (and therefore
also the resulting pressure on the boundary) is clearly unphysical,
while the two values within the range exemplify feasible
curvatures. We note that the value $\cell = 1$ is not sensitive to the
small changes in geometry resulting from a single migration event, and
that the population boundaries in such cases become more diffusive in
our experience. We set $\cell = 0.1$ for all uses of
\eqref{eq:elliptic_projection} in the experiments presented herein, a
choice which we have found works well across various examples.

Finally, to practically impose the pressure drop constraint given by
\eqref{eq:young-laplace} on the pressure equation, i.e.,
\eqref{eq:laplace_quantities} for $m = 0$, we extend the discretized
system of equations using Lagrange multipliers. When solving
\eqref{eq:laplace_quantities} using finite elements we generally
arrive at linear equations of the form $A\xi = b$, where $A$ is the
stiffness matrix and $b$ the load vector. Pressure drops defined by
\eqref{eq:young-laplace} against the ambient pressure are trivially
imposed in the system of equations as Dirichlet conditions given the
curvature $\hat{C}$. To
impose the pressure drops between different cell types, however, we
instead add the corresponding Lagrange multipliers to the system,
obtaining
\begin{align}
  \begin{split}
    \label{eq:Lagrange_multipliers}
    A\xi + \sum_{n=1}^N L_n \lambda_n &= b, \\
    L_n^T \xi \coloneqq \xi_i - \xi_j &= \sigma_{kl}\hat{C} \eqqcolon
    c_n,
  \end{split}
\end{align}
for the $n$th pair of interface nodes $\Omega_i$ and $\Omega_j$,
containing cells of type $k$ and $l$, respectively, where $L_n$ is a
vector the same size as $b$, and $N$ is the total number of
constraints. To solve \eqref{eq:Lagrange_multipliers} then amounts to
solving the extended system of equations
\begin{align}
    \begin{split}
    &\begin{bmatrix}
    A & L\\
    L^T & 0
    \end{bmatrix}
    \begin{bmatrix}
    \xi\\
    \lambda
    \end{bmatrix}
    =
    \begin{bmatrix}
    b\\
    c
    \end{bmatrix}, \\
    &L \coloneqq [L_1, L_2, ..., L_N], \\
    c \coloneqq [c_1,c_2,& ..., c_N]^T, \quad \lambda \coloneqq 
    [\lambda_1,\lambda_2, ..., \lambda_N]^T.\\
    \end{split}
\end{align}
We return to how these conditions are currently implemented over the
geometry of the framework voxels in the next section.

\subsection{Implementation details}
\label{sec:implementation_details}


In this section we assume some familiarity with stochastic simulations
and continuous-time Markov chains (CTMCs). We denote Markovian events
by $S \rightarrow S'$ which is to be understood that a state $S$ is
transformed into another state $S'$ after an exponentially distributed
waiting time with intensity $\lambda = \lambda(S)$. We refer the
reader to the monographs \cite{VanKampen, Gardiner} for more details
on the topic.

The DLCM solver discretizes the underlying space continuous physics as
detailed in \S\ref{sec:framework_formalism} and treats the rates of
change as propensities in CTMCs, with subsequent events sampled by
SSAs. One CTMC is defined for the external layer and another one for
the internal layer, cf.~\figref{2layer}.  To this end, we formally
define the mapping from continuous to discrete as
$u_i = u(x_i)\vert \Omega_i\vert$ (cell count), $v_i = v(x_i)$
(pointwise microphysical quantity), and
$w_i = w(x_i)\vert\Omega_i\vert$ (cellular internal state), where
$x_i$ is the midpoint of the voxel $\Omega_i$ and $\vert\Omega_i\vert$
the voxel volume. Subsequently, we use these discrete variables to
define the CTMCs, keeping in mind that they stem from continuous field
variables.

The DLCM framework handles cell migration as discrete events between
voxels, and the flux \eqref{eq:migration_physics} is discretized and
treated as an event propensity. In general, migration of one cell from
voxel $\Omega_i$ to $\Omega_j$ is denoted as an event
\begin{align}
  \label{eq:rates_migration}
  u_{i} \overset{I_{ij}}{\longrightarrow} u_{j},
\end{align}
where $I_{ij}$ is the flux between the two voxels. Clearly, the
notation \eqref{eq:rates_migration} is shorthand for the change of
cell number in the two neighboring voxels, with the cell's internal
state $w_i$ understood as moving along with it. Considering each term
in \eqref{eq:migration_physics} separately, the flux is discretized by
assuming that the migration potential gradient is constant across the
two voxels' shared edge and integrating over this edge, obtaining
\begin{align}
  \label{eq:rates_migration_explicit}
  I_{ij} = D(u_i, u_j; \; v_i, w_i) \times \frac{e_{ij}}{d_{ij}}
  \left[f(u_j, v_j, w_j)-f(u_i, v_i, w_i)\right],
\end{align}
with $e_{ij}$ being shared boundary length and $d_{ij}$ distance
between voxel centers (cf.~\figref{2layer}). Negative flux is
ignored. Although originally motivated from purely physical arguments
\cite{engblom2018scalable}, we note here that
\eqref{eq:rates_migration_explicit} may be regarded as a semi-discrete
\emph{upwind finite volume} scheme for the migration event
intensities. We return to this observation in
\S\ref{sec:solver_complexity} where we indicate an associated error
bound. Importantly, the scaling parameter of each migration potential
$D^{(n)}$ may depend on the number of cells in both voxels $i$ and $j$
which allows, for example, the implementation of spatial exclusion
mechanics of the cells in which migration to occupied voxels is
forbidden. Additionally, cell movement obeys a first-in, first-out
principle at each voxel, where the first cell to enter is also the
first to exit. This property ensures, for example, that one cell does
not carry information across the population by unphysically passing
through it.

The framework implements two specific restrictions to the migration
events: movements leading to $u_j>2$ are forbidden, as are movements
between voxels containing cells of different type whenever these
exhibit surface tension against one another. Both restrictions are
explicit design choices since they simplify the implementation and
improve the simulation efficiency considerably. The restriction
$u_j\leq C$ generally implies a spatial exclusion property for the
cells by which certain movements are impeded. The specific restriction
$C=2$ avoids a combinatorial growth of events that need to be defined
either implicitly or explicitly by the modeler. Finally, the
restriction concerning cell types expressing different surface tension
is necessary when using the same mesh for the pressure as for the cell
population. After applying the pressure drop \eqref{eq:young-laplace}
between cells of different type we cannot represent the discrete
pressure gradient in an upwind sense for migration of a cell into a
voxel containing a cell of different type: the pressure value in the
upwind voxel belongs to the other population. In the current
implementation, we thus exclude such events. Notably, neither of these
restrictions are an issue to expressing the cell sorting and
reorganization benchmarks as demonstrated in \S\ref{sec:cellsorting}
and \S\ref{sec:chemotaxis}.

The reaction events of the framework's external layer
\eqref{eq:external_layer_physics} are handled similarly to migration,
but occur locally in each voxel, i.e.,
\begin{align}
  \label{eq:rates_reaction}
  u_i \xrightarrow[]{\omega_r(u_i, v_i, w_i)} u_i + \Nstoich_r,
\end{align}
where $\omega_r$ is the propensity of reaction $r$ in $\omega$, and
$\Nstoich_r$ is the $r$th column of the stoichiometric matrix
$\Nstoich$ (cf.~\eqref{eq:external_layer_physics}). The external
layer reaction events \eqref{eq:rates_reaction} together with
\eqref{eq:rates_migration_explicit} define a CTMC from which the
solver samples the next event and the associated waiting time using an
exact SSA.

The internal layer of the solver handles the dynamics of the internal
cell states $w$ in \eqref{eq:internal_layer_physics}, and for each
cell independently,
\begin{align}
  \label{eq:rates_internal}
  w_i \xrightarrow[]{\nu_r(u_i, v_i, w_i)}  w_i + \Sstoich_r,
\end{align}
with $\nu_r$ the $r$th propensity in $\nu$ and $\Sstoich_r$ the
corresponding column of $\Sstoich$
(cf.~\eqref{eq:internal_layer_physics}). The independence between
cells can be relaxed to account for cell-cell signaling processes as
follows. Paracrine signaling (short-range diffusion of signaling
molecules) is included as a standard diffusion process between volumes
in the URDME setting; for contact-dependent juxtacrine signaling,
ligand-receptor binding can be modeled as the average of the
neighboring states leading to the production of a molecule in a cell,
e.g., setting $\nu_r \propto \sum_j w_j$ where $j$ runs over all
neighboring cells. For efficiency, we consider such incoming
juxtacrine signals to be approximately constant during some time
interval defined by the modeler. The internal events
\eqref{eq:rates_internal} occur between the Markov event times of the
external layer during which both $u_i$ and $v_i$ are assumed to be
constant, i.e., the two layers rely on separate Markov chains. This
decoupling improves the solver efficiency by avoiding solves for
\eqref{eq:laplace_quantities} after each internal state event and is
fundamentally made possible by the underlying scale separation
assumption of the two layers.

\algref{dlcm} summarizes the solver algorithm discussed thus far.

{\centering
  \begin{minipage}{1\linewidth}
    \begin{algorithm}[H]
      \if\submit1
      \doublespacing
      \fi
      \caption{DLCM solver.}\label{alg:dlcm}
      \begin{algorithmic}
        \State \textit{Initialize:} Given the mesh of voxels 
        $\Omega_i$ for $i = 1, 2, ..., \Nvox$, factorize the
        Laplacian and curvature operators. Populate the grid with
        cells, $u_i \in \{0,1,2\}$, each with local cell data, $w_i$
        (phenotype, molecular species, etc.). Initialize global
        quantities $v^{(m)}$, $m = 0, 1, \ldots M$.
        \While{$t < T$} \Comment{External layer}
        \State Calculate population curvature using
        \algref{elliptic_proj}.
        \State Update global quantities $v$ according to
        \eqref{eq:laplace_quantities} and \eqref{eq:young-laplace} for
        the pressure.
        \State Calculate reaction rates ($\omega_r$) and movement
        rates \eqref{eq:rates_migration_explicit}.
        \State Sample the event and the waiting time $dt$ by the
        Gillespie algorithm.
        \State Set $\tau \gets t$.
        \While{$\tau < t+dt$} \Comment{Update internal layer during $dt$}
        \If{internal states are continuous}
        \State Update synchronously with appropriate numerical scheme
        of time step $d\tau$.
        \Else
        \State Sample waiting time $d\tau$.
        \If{$\tau + d\tau > t + dt$} \Comment{Reject event}
        \State \textbf{break}
        \EndIf
        \State Sample event and update using the URDME SSA-solver.
        \EndIf
        \State $\tau \gets \tau + d\tau$.
        \EndWhile
        \State Set $t \gets t+dt$ and execute the external layer event
        sampled previously.
        \EndWhile
      \end{algorithmic}
    \end{algorithm}
  \end{minipage}
}

The solver has capacity for both continuous and discrete internal
states, respectively: the former are updated by an explicit first
order time stepping scheme in user-defined time intervals, and the
discrete states are updated using an internal CTMC using Gillespie's
direct method \cite{gillespie1976general} with URDME's built-in SSA
solver.

\begin{figure}[ht!]
  \centering
  \includegraphics[]{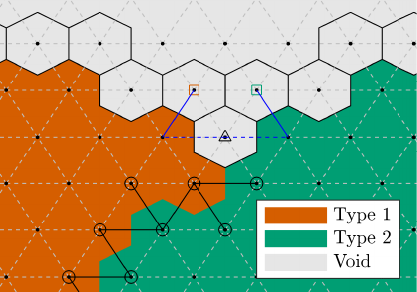}
  \caption{
    \if\submit1
    \doublespacing
    \fi
    \textbf{Pressure discontinuities on the grid}. Schematic of how
    the pressure discontinuity in \eqref{eq:young-laplace} is
    implemented on a finite element mesh (dashed grey) when it
    coincides with the cellular grid (hexagons). The figure shows
    cells of two different type (red and green) and empty voxels
    (grey). There exist two distinct situations for imposing a
    pressure discontinuity: between different cells or between a cell
    and an empty voxel. The former is implemented as a pressure
    difference between the adjacent nodes using
    \eqref{eq:Lagrange_multipliers}, here indicated by the black lines
    between circles. The latter condition, however, is set directly on
    the empty voxel adjacent to the cell (indicated here by the red
    and green squares). For the case when an empty voxel neighbors two
    cells of different type (triangle), a compromise must be made
    about which condition to apply.}
  \label{fig:curvatureBC}
\end{figure}

Finally, we describe the methods used to attain the micro-environment
quantities defined by \eqref{eq:laplace_quantities} before each
event. The full numerical pipeline reads: we solve for $v$ using FEM
with piecewise linear basis functions over a valid triangulation of
the mesh, then use the semi-discrete FV scheme
\eqref{eq:rates_migration_explicit} with the discrete pressure
$v^{(0)}$ to get the migration rates across the voxel edges (see
\figref{curvatureBC} for how the FEM mesh underlies the geometry of
voxels in a hexagonal mesh).  While there are many highly efficient
numerical methods to invert the Laplacian to solve for $v^{(m)}$, in
the numerical experiment reported here we simply use a direct
LU-decomposition of a finite element discretization over
$\Omegacomp$. We use the same approach for the additional operators
required for the elliptic projection of the curvature in
\algref{elliptic_proj}, as well as other custom operators highlighted
in \S\ref{sec:chemotaxis}. Potential improvements here are discussed
in \S\ref{sec:discussion}. Notably, in the current implementation the
mesh of the micro-environment quantities and the cells coincide, which
impacts how we impose the pressure drops in
\eqref{eq:young-laplace}. To maintain a representation of the upwind
pressure gradient out of a population, the pressure drop against the
environment is set directly on each empty adjacent voxel. We assume
that the ambient pressure is zero such that the cellular pressure due to
surface tension is known at that empty voxel, as indicated in
\figref{curvatureBC}. Thus, when an empty voxel neighbors two or more
cells of different type, we must choose which cell's condition applies for
this mesh configuration, and the current implementation selects the first
cell type out of all neighboring ones. We observe a consequence of the
resulting reduction in accuracy in \S\ref{sec:cellsorting}.

To summarize, the DLCM solver handles population-level events, such as
migration, switching phenotype, proliferation, and death in a CTMC
according to \eqref{eq:rates_migration} and \eqref{eq:rates_reaction},
respectively. Each cell can further be equipped with an internal state
that evolves in an internal CTMC according to
\eqref{eq:rates_internal}. Both CTMCs are sampled from using two
distinct SSA solvers. System micro-environment quantities are modeled
by Laplace equilibria \eqref{eq:laplace_quantities} that may affect
the reaction and migration rates of the cell and its internal state
dynamics. Population pressure may be constrained by the population
geometry through surface tension between cells of different type
according to \eqref{eq:young-laplace}, estimating the curvature
through elliptic projection using \algref{elliptic_proj}.  The DLCM
core simulation routines are currently implemented fully transparently
in Matlab. The solver objects and their relations are summarized in
\tabref{solver_input}.

\begin{table}[H]
  \begin{threeparttable}
  \caption{DLCM solver objects and a summary of their function.}
  \if\submit1
  \doublespacing
  \fi
  \centering
  \begin{tabular}{p{1cm}p{13.7cm}}
    \hline
    Object & Function [unit] \\
    \hline
    $u_i$ & Cell number in voxel $\Omega_i$ $[\# cells]$ \\
    $w_i$ & Cell internal data of cells in voxel $\Omega_i$ $[\# species]$ \\
    $v$ & Micro-environment concentrations $[l^{-d}]$\tnote{ab} \\
    $\Nstoich$ & Stoichiometric matrix for the external cell events,
                   e.g., phenotype switching, proliferation, and death
                   $[\# cells]$  \\
    $\omega_r$ & Event propensities corresponding to $\Nstoich$ $[t^{-1}]$ \\
    $\Sstoich$ & Stoichiometric matrix of the internal cell state dynamics,
    such as protein or signaling molecule production $[\# species]$ \\
    $\nu_r$ & Event propensities corresponding to $\Sstoich$
              $[t^{-1}]$ \\
    $f^{(n)}$ & Functions of the micro-environment quantities $v^{(m)}$ whose
    spatial gradients drive cell migration $[\cdot]$\tnote{c} \\
    $D^{(n)}$ & Flux scaling factor for $\nabla f^{(n)}$. May depend
                on the number of cells in neighboring voxels to allow
                for spatial exclusion effects $[l^{2} t^{-1}\times
                \cdot ]$\tnote{c} \\
    $s^{(m)}$ & Source for $v^{(m)}$,
                cf.~\eqref{eq:laplace_quantities} $[l^{-d-2}]$ \\ 
    $r^{(m)}$ & Dirichlet boundary condition for $v^{(m)}$ $[l^{-d}]$ \\
    \hline
  \end{tabular}
  \begin{tablenotes}
    \item[a] The units $l$ and $t$ denote length and time, respectively.
    \item[b] $d$ is the number of spatial dimensions. 
    \item[c] The unit of $f^{(n)}$ depends on its formulation, and the
      inverse of this unit multiplied by $l^{2} t^{-1}$ comprises the
      unit of $D^{(n)}$.
  \end{tablenotes}
  \label{tab:solver_input}
  \end{threeparttable}
\end{table}

\subsection{Solver consistency and computational complexity}
\label{sec:solver_complexity}


To get a feeling for in what sense the framework is consistent with
known physics, we look at the case of purely pressure-driven
advection, or Darcy's law:
\begin{align}
  \label{eq:pressure-driven_model}
  \left. \begin{array}{rcl} \frac{\partial u}{\partial t}
           - \nabla\cdot(uD\nabla p) &=& 0 \\
           -\Delta p &=& f \\
         \end{array}\right\}
\end{align}
at $\Omegacomp$ and with $p = u = 0$ at $\partial \Omegacomp$ together
with some given initial data $u_0$. First, DLCM solves for $p_h$
defined by the variational formulation (cf.~\eqref{eq:inner_product})
\begin{equation}
  \label{eq:pressure_perturbed_source}
  (\nabla p_h,\nabla v) = (f_h,v) = (f,v),
  \quad \forall v \in V_h,
\end{equation}
using piecewise linear basis functions on the mesh discretizing
$\Omegacomp$. Here $f_h$ is the projection of $f$ onto $V_h$ with
$\lVert f_h-f\rVert = \Ordo(h^2)$ and where $h$ is the mesh size,
i.e., the diameter of a single cell. This can be viewed as a perturbed
source function of $p_h$, for which $\lVert p_h-p\rVert = \Ordo(h^2)$
\cite{larson2013finite}.

Given $p_h$ the framework samples the stochastic process, $U_h(t)$,
using the migration rates \eqref{eq:rates_migration_explicit}, which
we now aim to compare against the \emph{perturbed} solution $u^*(t)$
to \eqref{eq:pressure-driven_model} under the perturbation
$p\mapsto p_h$. Specifically,
\begin{itemize}
\item let $U_h(t)$ be sampled from the CTMC driven by the rates
  $U_i \xrightarrow[]{-U_i(D\nabla p_h)_{ij}} U_j$ given the initial
  data $U_h(0)$;
\item let $u_h(t)$ solve for $u^*$ on the same mesh using a
  semi-discrete upwind FV scheme equivalent to the DLCM migration
  rates \eqref{eq:rates_migration_explicit}.
\end{itemize}
By linearity of the model problem considered we have that
$u_h(t) = \Expect[U_h(t)]$ such that an error in $u_h$ can be
understood as a bias term of the DLCM solution. In
\S\ref{apx:proof_estimate} we show that the upwind scheme is
equivalent to a standard first order FV scheme for $u^*$, implying
that the bias in $U_h(t)$ is consistent with
\eqref{eq:pressure-driven_model} to first order in $h$. Thus, we
obtain an $\Ordo(h)$-biased solution to a problem involving an
$\Ordo(h^2)$-perturbed pressure, where $h$ is a measure of the length
scale of a single cell.

Finally, we briefly discuss an estimate for the solver computational
complexity for models where pressure-driven cell migration is the
dominating mechanism behind the flow of cells. Similar arguments may
be used to derive the complexity for other types of migration
potentials.
\begin{proposition}[\textit{DLCM solver complexity bound}]
  \label{prop:solver_complexity}
  Assume in a certain DLCM model $M$ that the number of cells are
  bounded by $\Ncells$ during a time interval $(0,T)$, and that the
  total external and internal event intensities for each cell are
  uniformly bounded as $\|\omega\|_1 \le \mathcal{W}$ and
  $\|\nu\|_1 \le \mathcal{V}$, respectively. Assume that cell
  migration is dominated by pressure-driven mechanics. Let
  $\Ccomplexity$ denote the computational complexity of the model $M$
  in the sense of arithmetic operations. Note that this is a
  stochastic variable, whose value depends on the sample.

  Then we have the following bound on the expected value:
  \begin{equation}
    \label{eq:solver_complexity}
    \Expect[\Ccomplexity] = \Ordo\bigg(T \Ncells \times \big( 
    \overbrace{1}^{\mathrm{migration}} +
    (\overbrace{\sigma_{\max}\Nvox^{1/d}}^{\mathrm{surfac
        e \ tension}} + 
    \overbrace{ 
      \mathcal{W}}^{\mathrm{ external}})\times 
    \overbrace{\Nvox}^{\mathrm{Laplace}}  + 
    \overbrace{\mathcal{V}}^{\mathrm{internal}} \big) \bigg),
  \end{equation}
  where $\sigma_{\max} \equiv \max_{kl}\sigma_{kl}$ and
  $d \in \{2,3\}$ is the number of spatial dimensions. The complexity
  for systems based on purely pressure-driven migration is thus
  optimal, scaling linearly with $T$ and $\Ncells$.
\end{proposition}

We formally derive Proposition~\ref{prop:solver_complexity} in
\S\ref{apx:proof_complexity}. Note that the assumptions made may be
violated for certain models or population configurations leading to a
different asymptotic behavior, e.g., when using a more involved
migration potential or custom micro-environment operators. On the
other hand, the analysis might also be overly pessimistic whenever the
number of cells with a non-zero migration rate, for example, might
rather be proportional to the number of boundary cells, i.e.,
$\Ncells^{1/d}$ instead of $\Ncells$. Also, the surface tension term
is derived assuming the worst possible bound for the curvature,
$C\propto h^{-1}$, and then for \emph{all} cell-cell interfaces.


\section{Results}
\label{sec:results}

We present here four example models handled by the DLCM solver
highlighting the diverse capacity of the framework. We consider a set
of models of similar character to widely cited models of cell
populations: the benchmark case of cell sorting, cell signaling and
patterning of a Hes1 signaling model, tumor growth, and a variety of
chemotaxis models. The execution time in the current implementation
for all experiments were on the order of ten minutes to an hour on a
standard laptop simulating up to about $10^3$ cells.  Most of the
simulation time is consumed by inverting the Laplace operator
governing the cellular pressure and for which highly optimized
numerical solvers do exist. However, in the first release of DLCM as a
solver in URDME, a direct solver using Matlab's sparse
LU-decomposition has been used for this purpose.

\subsection{Cell Sorting}
\label{sec:cellsorting}


A striking benchmark problem for discrete cell population models is
the problem of cell sorting \cite{osborne2017comparing}. Here, cells
spontaneously sort themselves by different cell phenotypes, a
phenomenon observed for biological cells. As mentioned in
\S\ref{sec:curvature_surfacetension} different hypotheses have
been proposed to explain the process, such as differences in cell-cell
adhesion or interfacial tension
\cite{tsai2022selforgreview}. Computational modeling has been
instrumental in elucidating the potential mechanisms involved
\cite{glazier1993simulation, brodland2015computational}.

We set up the model for cell sorting by considering two different cell
types placed in a random configuration. We assume Darcy's law for the
migration of cells,
\begin{align}
  \label{eq:cellsort_rates}
  u_i \xrightarrow[]{(-D\nabla p)_{ij}} u_j,
\end{align}
where $p$ is the pressure field defined as
\begin{align}
  \label{eq:cellsort_quants}
  \left. \begin{array}{rcl}
           -\Delta p &=& s(u) \\
          p(x^+) - p(x^-) &=& \sigma_{kl} C
                         \end{array} \right\}
\end{align}
with $x^{\pm}$ as in \eqref{eq:young-laplace}, and with the
Young-Laplace pressure drop both between cells of different type
($\sigma_{12}$) and towards the external medium ($\sigma_{10}$,
$\sigma_{20}$). Here $s(u) = 1$ only in voxels with $u_i > 1$,
representing pressure exerted in overcrowded spaces. We consider two
cases: $\sigma_{10} = \sigma_{20} < \sigma_{12}$ and
$\sigma_{12} < \sigma_{10} < \sigma_{20}$ (parameter details in
\tabref{parameters_cell_sorting}), denoted respectively the symmetric
and asymmetric cases.

To calibrate the timescale of the model, we consider that cells can
move roughly one cell diameter per minute, and we assume that the
model's cells do so under a unit pressure gradient. Then, from
\eqref{eq:rates_migration_explicit}, $e_{ij}D\times1=60$ hours$^{-1}$, which
implies $D = 4200 [(tf)^{-1}]$ with the timescale in hours.
\begin{table}[H]
 \begin{threeparttable}
 \caption{Parameters of the cell sorting experiments.}
  \if\submit1
  \doublespacing
  \fi
  \centering
  \begin{tabular}{ p{4cm}p{2.9cm}p{7.2cm}  }
    \hline
    Parameter & Value & Description\\
    \hline
    $[\sigma_{10}, \sigma_{20}, \sigma_{12}]$  & $[1, 1, 5] \times
    10^{-4}$ & Symmetric surface tension coefficients \\
    $[\sigma_{10}, \sigma_{20}, \sigma_{12}]$  & $[2, 4, 1] \times
    10^{-4}$ & Assymetric surface tension coefficients \\
    $[D(1,0), D(2,0), D(2,1)]$\tnote{a} & [1, 1, 25]$\times4200$ & Migration rate
                                                       scaling (both cases) \\
    \hline
  \end{tabular}
  \begin{tablenotes}
  \item[a] For brevity, we define
    $D(u_i,u_j) \equiv D(u_i,u_j; \; v_i,w_i)$. Except for the cases
    listed above, $D(m,n)$ is zero, however, we set $D(1,1) = 1$ for
    voxels where the pressure jump conditions are applied.
  \end{tablenotes}
  \label{tab:parameters_cell_sorting}
  \end{threeparttable}
\end{table}

\begin{figure}[H]
  \centering
  \includegraphics{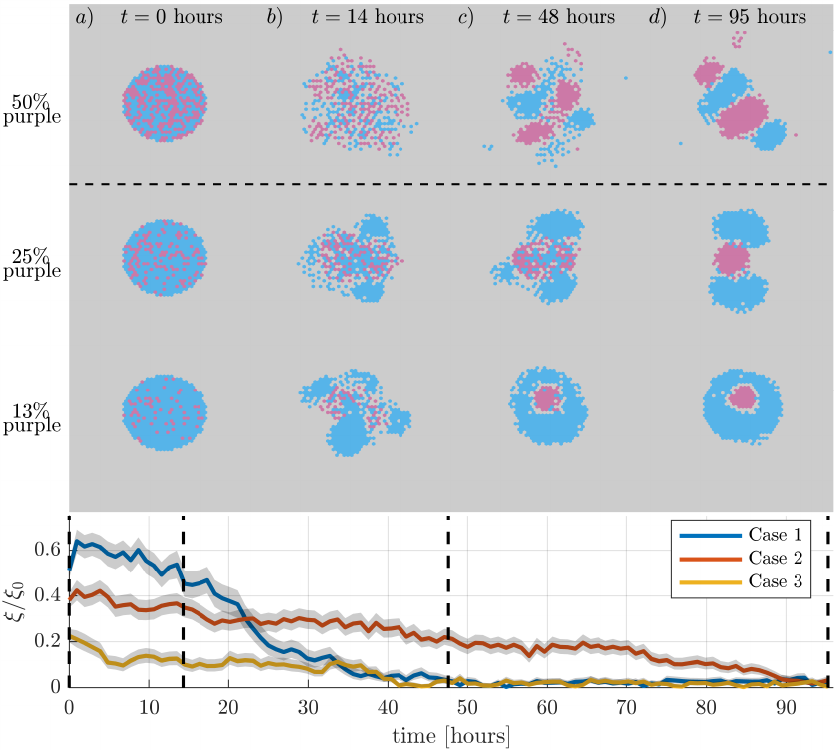}
  \caption{
    \if\submit1
    \doublespacing
    \fi
    \textbf{Cell sorting experiments.} Three examples of spontaneous
    sorting of cells of two different types from an initially random
    configuration \textit{(left)}, to a segregated population
    \textit{(right)}. The top row shows the symmetric experiment with
    an equal number of cells of both types. The two rows below show
    the asymmetric experiments, with 25\% and 12.5\% cells of type $2$
    (purple), respectively. Engulfment occurs when the second type is
    significantly outnumbered. The columns \textit{a)--d)} each
    correspond to a different time: $0$, $14$, $48$, and $95$ hours,
    respectively. The bottom figure shows the normalized fractional
    length over time for each of the three experiments, with grey
    indicating the 68\% confidence interval. The dashed lines show the
    four times of the above experiment snapshots.}
 \label{fig:cellsort}
\end{figure}

The resulting sorting is shown in \figref{cellsort} for both cases. We
note that the experiment for the symmetric case results in a cell
fragmentation akin to the overlapping spheres model in
\cite{osborne2017comparing}, but that we also achieve \emph{engulfing}
of one type into the other in the second experiment of the asymmetric
case, akin to the Vertex, Cellular Potts and Cellular Automata models
in \cite{osborne2017comparing}.

For each experiment we calculate the normalized fractional length,
defined similar to the associated metric in
\cite{osborne2017comparing}, but here normalized differently. Let
$\xi(t)$ be the total edge length between cells of different type and
$\xi_0(t)$ be the total edge length irrespective of type, both
evaluated at time $t$. Using a hexagonal grid, each discrete edge
length, $e_{ij}$, is equal and we can view the number of edges
$\xi/e_{ij}$ as a random variable of a binomial distribution
$\xi/e_{ij} \sim B(\xi_0/e_{ij},\phi)$, where
$\hat{\phi}(\xi)=\xi/\xi_0$ is an estimator of $\phi$. In
\figref{cellsort} we plot this estimator together with the 68\%
confidence interval. We see that the estimator $\hat{\phi}$ shrinks
for all experiments as expected; in fact, the fractional length in our
experiments shrinks closer to zero than it does in the experiments in
\cite{osborne2017comparing}. The smaller fractional length is
explained by the one voxel gap which for example is visible in case 3
in \figref{cellsort}, and which in turn can be explained as
follows. Recall from \S\ref{sec:implementation_details} that pressure
drops versus the ambient pressure are imposed on the empty voxel
neighbouring the cell; however, when an empty voxel
neighbours two or more cells with different surface tension against
it, a decision must be made over which condition to apply there. The
net effect is that a stationary distribution of cells is obtained
where a one voxel gap separates the two types. A more highly resolved
model would be one where a finer grid for the pressure could better
represent the upwind pressure gradient out from the population
boundaries, and would thus ensure a potentially better resolution of
these specific kinds of boundary regions. This case highlights a
specific modeling limitation of certain movements when using the same
grid for the pressure as for the cells.

\subsection{Cell Signaling}
\label{sec:cellsignaling}


The DLCM framework couples the dynamics of the internal cell states
with the outer level dynamics (migration, birth, death, etc.)~by using
an internal SSA solver. We show this capability through a model of
cell-cell signaling in a growing population, using a model of
Hes1-Notch dynamics recently studied in the RDME setting
\cite{menz2025modelling}. Using their notation, we let [$D$, $N$, $M$,
$P$, $n$] represent the number of species of Dll1, Notch, Hes1 mRNA,
Hes1 protein, and Ngn2, respectively (omitting the cell subscript $i$
for brevity). In essence, it is an extension of the Delta-Notch model
\cite{collier1996pattern} which includes the Hes1 negative feedback
loop. A sufficient model to achieve the proper dynamics is attained by
translating their ODE directly to Markov events as
\begin{align}
  \label{eq:DeltaNotch_rates}
  \left. \begin{array}{rclrcl}
            n & \xrightarrow[]{\alpha_D n}  & n + D, &
            D & \xrightarrow[]{\mu_D D}  & \emptyset,  \\
            \emptyset & \xrightarrow[]{\alpha_N \langle D_{in} \rangle}  & N, &
           N & \xrightarrow[]{\mu_N N}  & \emptyset, \\
           N & \xrightarrow[]{\alpha_M N f(P)}& N + M &
           M & \xrightarrow[]{\mu_M} & \emptyset \\
           M & \xrightarrow[]{\alpha_P M}  & M + P, &
           P & \xrightarrow[]{\mu_P P}  & \emptyset, \\
           \emptyset & \xrightarrow[]{\alpha_n g(P)}  & n, &
           n & \xrightarrow[]{\mu_n n}  & \emptyset, 
          \end{array} \right\}
\end{align}
where
\begin{align}
  \left. \begin{array}{rcl}
            f(x) &=& 1/(1 + (x/(K_MV))^k) \\
            g(x) &=& 1/(1 + (x/(K_nV))^h) \\
            \langle D_{in} \rangle &=& 
            |\mathcal{N}|^{-1}\sum_{i\in\mathcal{N}}D_i
          \end{array} \right\}
\end{align}
where $V$ is the cell volume, $\mathcal{N}$ is the set of indices of
the neighboring cells, and $|\mathcal{N}|$ is the number of
neighbors. The initial expression levels are generated as in
\cite{menz2025modelling} with a 5\% perturbation (lognormally
distributed) from a fixed value in every cell. The cells migrate
by Darcy's law and the pressure field \eqref{eq:cellsort_quants} but
with zero $\sigma_{10}$ (there is only one cell type in this
model). For each cell, we include a proliferation rate
$\mu_{\mathrm{prol}}$ if it is within a radius of $0.2$ within the
origin. The initial population is also a circle of radius $0.2$, or
$0.45$ when $\mu_{\mathrm{prol}} = 0$.  The proliferation rate is
turned off after five full cell proliferation cycles (i.e., after
$5\mu_{\mathrm{prol}}^{-1}$ time units), such that it grows to the
size of the static population, and each simulation continues to run
to the same fixed time. The framework is capable of handling
both discrete and continuous internal states, and we run simulations
for both cases where the latter considers the concentrations of
respective molecule obeying the original ODE. The system volume $V$ is
set to $50\mu$m$^2$, although a range of other values still yield
system patterning \cite{menz2025modelling}.

\begin{figure}[H]
  \centering
  \includegraphics{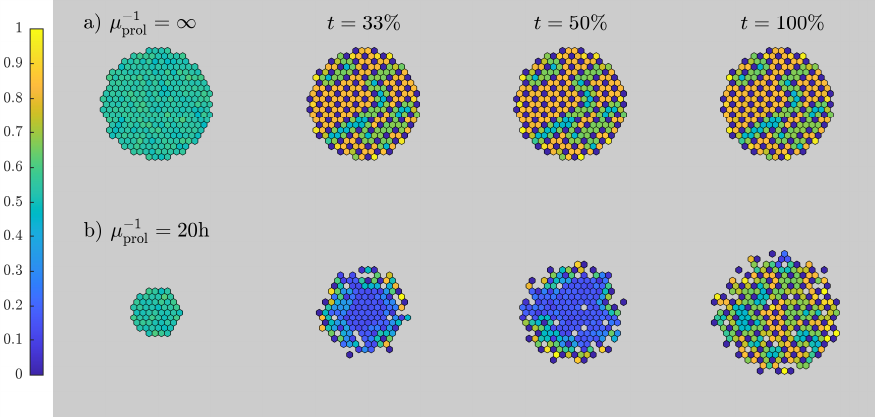}
  \caption{
    \if\submit1
    \doublespacing
    \fi
    \textbf{Cell signaling experiments (Discrete)}. Outcome of the
    Hes1 model with cell proliferation, figures a) and b) showing the
    number of Hes1 proteins $P$ (normalized to the range $[0,1]$) for
    two different values of the proliferation rate,
    $\mu_{\mathrm{prol}} = 0$ and $1/(20\times60)$, respectively. Both
    simulations are run until $120$ hours, and we show the outcome of
    both experiments at $0$, $33$, $50$, and $100\%$ of the simulation
    time.}
  \label{fig:signaling_discrete}
\end{figure}

We first show the outcome of the experiments with discrete internal
states in \figref{signaling_discrete} for both a static population and
one where the cells divide once every twenty hours on average. In the
static case, we achieve the typical salt-and-pepper patterning. The
growing population ultimately yields a similar patterning after the
proliferation has stopped, albeit with notably less regularity since
the patterning that starts from the edges is perturbed by the growth
of the population.

\begin{figure}[H]
  \centering
  \includegraphics{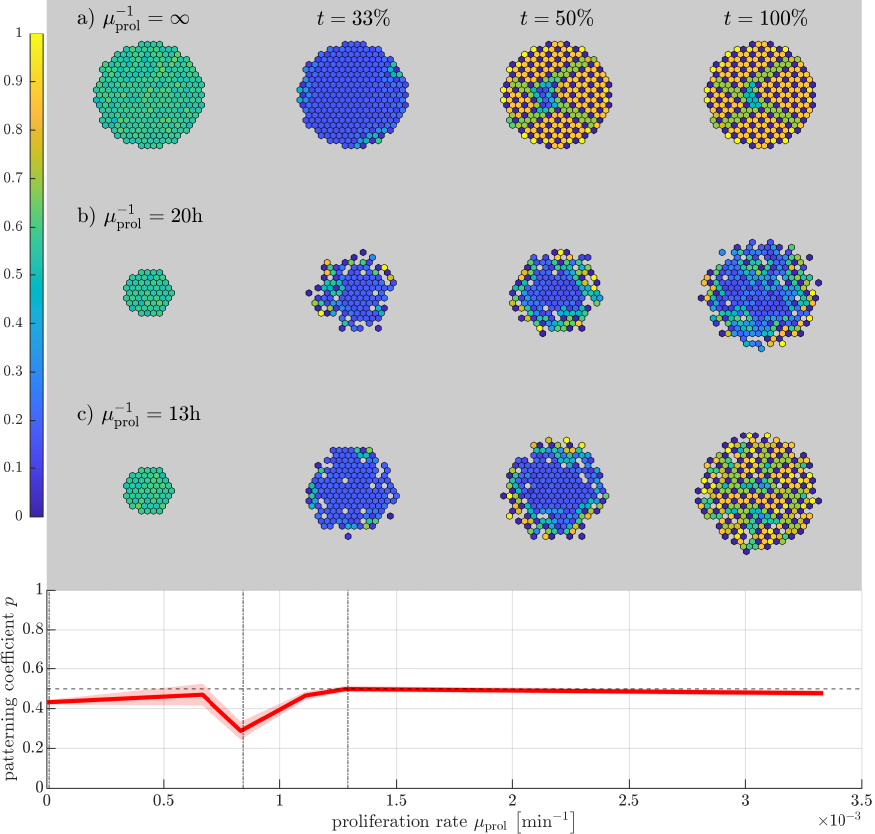}
  \caption{
    \if\submit1
    \doublespacing
    \fi
    \textbf{Cell signaling experiments (Continuous)}. Hes1 model with
    cell proliferation, figures a) and b) showing the normalized
    concentration of Hes1 protein $P$ for three different values of
    the proliferation rate $\mu_{\mathrm{prol}}$, the smallest,
    largest, and an intermediate value, respectively, from the range
    of values tested. The bottom figure shows the mean degree of
    patterning and standard deviation of 10 sample runs by the end of
    the simulation for a range of values of the proliferation
    rate. The static case is run for 84 hours, and we show the outcome
    of all experiments at $0$, $33$, $50$, and $100\%$ of the
    simulation time.}
  \label{fig:signaling_continuous}
\end{figure}

To better understand the robustness of this particular patterning
against a dynamic population, we simulate the Hes1 model over a range
of values of the proliferation rate $\mu_\mathrm{prol}$ using
continuous internal states. We run each simulation for $84$ hours such
that the static population just achieves a stationary internal state
to properly compare how the patterning is affected by different growth
rates.  During each simulation, we measure the patterning by counting
the number of neighbour couplings between two cells with high
expression levels as in \cite{menz2025modelling} (and where $p=0.5$
corresponds to perfect checkerboard patterning). To capture the
variability across outcomes, we run 10 sample simulations per value of
the proliferation rate, and similar to the fractional length metric in
\S\ref{sec:cellsorting}, we can treat the counts as independent
Bernoulli trials to assign a confidence interval for our patterning
estimates. \figref{signaling_continuous} shows one of the outcomes for
the experiments with no proliferation, slow and fast proliferation,
respectively.  We observe that the rate of patterning drops around an
intermediate range of values; then there is a shift towards perfect
patterning for proliferation times faster than $20$h with a steady
patterning close to $0.5$ for $13$h and below. Patterning is destroyed
for values close to $20$h just when the population is growing fast
enough but does not manage to reach the target size to stop growing
before simulation end; however, when growing even faster the
population does not loose the ability to generate a pattern since it
does so anyways after the growth ceases.

\begin{table}[H]
 \begin{threeparttable}
 \caption{Parameters of the Hes1 cell signaling experiments.}
  \if\submit1
  \doublespacing
  \fi
  \centering
  \begin{tabular}{ p{4cm}p{6.1cm}p{4cm} }
    \hline
    Parameter & Value & Description\\
    \hline
    $\mu_{\mathrm{prol}}$ & in range $0$ to $1/300$ minutes$^{-1}$ & Proliferation rate \\
     $[\alpha_D$, $\alpha_N$, $\alpha_M]$  & $[0.018, 6.0, 0.017]$ & Growth parameters \cite{menz2025modelling} \\
     $[\alpha_P$, $\alpha_n]$  & $[0.14, 0.0049]$ & Growth parameters \cite{menz2025modelling} \\
     $[\mu_D$, $\mu_N$, $\mu_M$, $\mu_P$, $\mu_n]$ & $[250$, $200$, $24.1$, $22.3$, $21.9 ]^{-1}\times \log(2)$ & Decay parameters \cite{menz2025modelling} \\
    $[K_M$, $k$, $K_n$, $h]$ & $[0.050, 0.030, 1, 4]$ & \cite{menz2025modelling}\\
    $D(u_i,u_j)$\tnote{a} & $100\times(u_j<u_i)$\tnote{b} & Migration rate scaling\\
    \hline
  \end{tabular}
  \begin{tablenotes}
    \item[a] $D(u_i,u_j) = D(u_i,u_j; \; v_i,w_i)$.
    \item[b] Boolean interpretation,  i.e.,  0 (false) or 1 (true).
  \end{tablenotes}
  \label{tab:parameters_delta_notch}
  \end{threeparttable}
\end{table}

\subsection{Tumor Growth}
\label{sec:tumorgrowth}


Cell population models that represent cells as discrete agents or
collections of agents have been used extensively to model various
aspects of tumor growth \cite{szabo2013cellular, phillips2020hybrid,
  lima2021bayesian}. We highlight here one of the simplest possible
models of avascular tumor growth that reproduce known qualitative
behavior \cite{engblom2018scalable}. Chiefly, aggressive tumor cells
proliferate if they have enough oxygen to do so, and become necrotic
if the have insufficient oxygen to stay alive. Live cells consume
oxygen, necrotic cells do not and they also degrade over time. Cells
in crowded regions exert pressure on surrounding cells that they relax
by migrating away from high-pressure regions. The tumor cells also
adhere to each other to a certain extent.

This model is conveniently translated into the DLCM framework as
follows. Let $u_i$ denote the number of live tumor cells in voxel
$\Omega_i$, $d_i$ the number of degrading necrotic tumor cells, $p$
the cellular pressure field, and $c$ the nutrient field. Darcy's law
governs the cell migration through the surrounding extra-cellular
matrix (ECM) due to the cellular pressure. We have four
reaction-transport events defined as
\begin{align}
  \label{eq:tumor_rates}
  \left. \begin{array}{rclrcl}
           u_i& \xrightarrow[]{\mu_{\mathrm{prol}}(c > \kappa_{\mathrm{prol}})}  &u_i + u_i, 
           & d_i& \xrightarrow[]{\mu_{\mathrm{deg}}}  &\emptyset \\
           u_i& \xrightarrow[]{\mu_{\mathrm{die}}(c < \kappa_{\mathrm{prol}})} &d_i, 
           & u_i& \xrightarrow[]{(-D\nabla p)_{ij}} &u_j
         \end{array} \right\}
\end{align}
where $(c\lessgtr \cdot)$ is interpreted as a boolean condition and where
$D$ is zero if the migration is to a voxel with $u_j=2$. Movement between
voxels with one cell each is generally forbidden, except from voxels
on the population boundaries to ensure that the surface tension functions
properly \cite[Appendix~B]{blom2024morphological}. The pressure and
nutrient field are modeled as
\begin{align}
  \label{eq:tumor_quants}
  \left. \begin{array}{rclrcl}
           -\Delta p &=& s(u), &
                                 [p(x)] &=& \sigma_{kl} C, \\
           -\Delta c &=& -\lambda u, &
                                       c &=& 1, \quad \text{at } \partial \Omegacomp,
         \end{array} \right\}
\end{align}
where $[p(x)]$ is the pressure jump at $x$,
$[p(x)] \equiv p(x^{+}) - p(x^{-})$ with $x^{\pm}$ as in
\eqref{eq:young-laplace}, and where $s(u) = 1$ only in voxels with
$u_i > 1$, representing pressure exerted in overcrowded spaces as
discussed in \S\ref{sec:cellsorting}.

\begin{figure}[ht]
  \centering
  \includegraphics{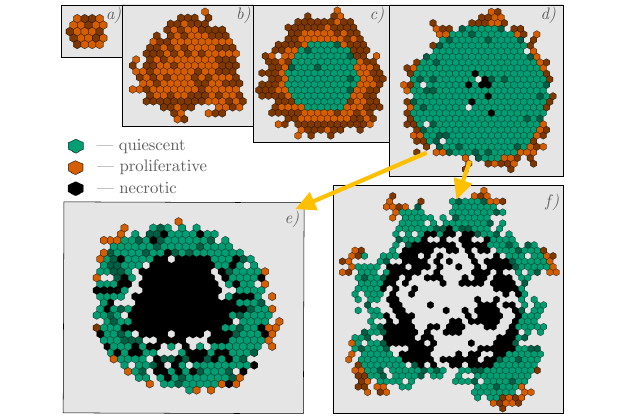}
  \caption{
    \if\submit1
    \doublespacing
    \fi
    \textbf{Avascular tumor growth experiments}. Red cells have
    sufficient oxygen and are proliferating, green cells are quiescent
    due to lack of oxygen, and black cells are necrotic
    (cf.~\eqref{eq:tumor_rates}), with darker shades indicating that a
    voxel contains two cells. Figures \textit{a)--d)} show the tumor
    growth in time, and figures \textit{e)}, \textit{f)}, highlight
    the morphological discrepancy that emerges due to high and low
    surface tension coefficient $\sigma$, respectively.}
\label{fig:tumor}
\end{figure}

Using the parameters in \tabref{parameters_tumor_growth} we get the
growth patterns shown in \figref{tumor}, where different morphologies
emerge due to differences in surface tension. The different
morphologies can be explained by analysis of a corresponding PDE
model, facilitated by the basis of the framework in time and space
continuous physics \cite{blom2024morphological}. In fact, in this
paper, the net volumetric growth rates of the tumor could be
accurately predicted from the PDE and compared well with the DLCM
model.

\begin{table}[H]
 \begin{threeparttable}
 \caption{Parameters of the tumor growth experiments.}
  \if\submit1
     \doublespacing
  \fi
  \centering
  \begin{tabular}{ p{4cm}p{2.9cm}p{7.2cm} }
    \hline
    Parameter & Value & Description\\
    \hline
    [$\mu_{\mathrm{prol}}, \mu_{\mathrm{die}}, \mu_{\mathrm{deg}}$] &
    $[10,10,1] \times 10^{-4}$ & Proliferation, death, and degradation rates \\
    $\kappa_{\mathrm{prol}}, \kappa_{\mathrm{die}}$  & $[0.955,0.945]$ &  Oxygen
    thresholds: proliferation and death \\
    $\lambda$  & 1 & Oxygen consumption \\
    $[D(1,0), D(2,0), D(2,1)]$\tnote{a} & $[1, 1, 25]$ & Migration rate scaling \\
    $[\sigma_{10} ,\sigma_{20}, \sigma_{21}]$  & $[10^{-3}, 10^{-3}, 0]$ & Surface
    tension coefficients \\
    \hline
  \end{tabular}
  \begin{tablenotes}
  \item[a] As before, $D(u_i,u_j) = D(u_i,u_j; \;
    v_i,w_i)$. Except for the cases listed above, $D(m,n)$ is zero,
    however, we set $D(1,1) = 1$ for voxels where the pressure jump
    conditions are applied.
  \end{tablenotes}
  \label{tab:parameters_tumor_growth}
  \end{threeparttable}
\end{table}

\subsection{Chemotaxis}
\label{sec:chemotaxis}


Chemical signals in the cell micro-environment can detail the presence
and quality of nutrients, toxic compounds, and other cells. Cells use
this information to, among other things, direct their movement through
the process of chemotaxis---a key mechanism in, e.g., bacterial
function, embryology, immune response, and cancer development
\cite{keegstra2022ecological, scarpa2016collective,
  roussos2011chemotaxis}. An early mathematical formulation of this
process is the Keller-Segel model of chemotaxis, initially used to
model the aggregation of slime molds \cite{keller1970initiation},
later branching into myriad variants for different applications in
mathematical biology---see \cite{arumugam2021keller} for a review on
the mathematical and numerical analysis of such models.

We highlight the framework's capacity to handle various forms of
chemotaxis by a few examples. The general Keller-Segel model considers
cell diffusion due to random motion balancing out the aggregation of
the chemotaxis, but we simulate models with both diffusive and
pressure-driven migration. The specifics of the chemotaxis and the
chemical signaling in all these models -- except for one -- are taken
from the comprehensive list of models in
\cite{arumugam2021keller}. Chiefly, model 1 uses pressure-driven
migration and chemotaxis upwards a constant chemical gradient, whereas
model 2 uses diffusive motion instead of Darcy's law.  Model 3 is
similar to model 1 but with chemotaxis \emph{downwards} the gradient
of a chemical that cells also consume, causing an aggregating
effect. Model 4 instead combines the pressure and the chemical into
one field whose gradient guides cell motion (using the chemotaxis
model in \cite{engblom2018scalable}), and, finally, the cells in model
5 emit a chemical that attracts other cells.

Specifically, the models are defined as follows. For the initial cell
population we set $u_i = 1$ within a circle of radius $0.25$ (but set
$u_i=2$ for the fourth model as the signal only modifies an existing
pressure). The cell state updates are defined purely by migration
events according to
\begin{eqnarray}
  \label{eq:chemotaxis_rates}
  u_i& \xrightarrow[]{\omega_m(u,p,s)} &u_j,
\end{eqnarray}
for models $m = 1, 2, ..., 5$, where $s$ is the chemical substance and
the model specific movement rates are
\begin{equation}
  \label{eq:chemotaxis_models}
  \omega_{m} = -G_{m}(\nabla u)_{ij}  -D_{m}(\nabla p)_{ij} - \chi_{m}
  (\nabla s)_{ij},
\end{equation}
where the parameters control the nature of the migration. Here,
$\chi_m$ is cell chemotaxis sensitivity and $G_m$ the diffusion rate
scaling for model $m$. The pressure and chemical signaling fields are
modeled as
\begin{align}
  \label{eq:chemotaxis_quants}
  \left. \begin{array}{rclrcl}
           -\Delta p &=& p_m(u), & p &=& 0, \quad \text{at } \partial
                                         \Omegapop  \\
           -\Delta s &=& s_m(u) , & s &=& b_m, \quad \text{at }
                                          \partial \Omegacomp \\
         \end{array} \right\}
\end{align}
where $\Omegapop$ is the population boundary and $p_m(u)$ defines
pressure sources for overcrowded voxels for all models except for the
fourth, as explained below. The signal source $s_m(u)$ models that
cells \emph{consume} signals for model three and \emph{emits} signals
for model five (and a static signaling field for the rest of the
models). The external source $b_m$ of the signal is equal to one on
the left boundary for models one, two, and four (but on the
\emph{right} boundary for model three) and zero everywhere else,
yielding a constant gradient across the domain in the absence of
consumption. The precise nature of these functions is presented in
\tabref{parameters_chemotaxis}.

\begin{table}[H]
  \begin{threeparttable}
    \caption{Parameters of the chemotaxis experiments.}
    \if\submit1
    \doublespacing
    \fi
    \centering
    \begin{tabular}{ p{3cm}p{5.9cm}p{5.2cm} }
     \hline
    Parameter & Value & Description\\
    \hline
    $[G_1, ..., G_5]$ & $[0, 1, 0, 0, 0]\times$scale \tnote{d}& Diffusion rate scaling\\
    $[D_1, ..., D_5]$ & $[1, 1, 1, 1, 1]\times$scale \tnote{a,d} & Darcy's law coefficient \\
    $[\chi_1, ..., \chi_5]$ & $[1, 1, -1, 0, \chi_5]\times$scale \tnote{b,d} & Chemotactic sensitivity \\
    $\chi$ & $-2\times10^{4}$ & Signal-pressure coupling \\
    $[p_1, ..., p_5]$ & $[1, 1, 1, 1, 10] \times \delta_{u,2}$\tnote{c} & Pressure
    sources, \eqref{eq:chemotaxis_quants} \\
    $[s_1, ..., s_5]$ & $[0, 0, -u, 0, u]$ & Signal sources,
    \eqref{eq:chemotaxis_quants}\\
    $[b_1, ..., b_5]$ & $[\delta_{x,-1}, \delta_{x,-1}, \delta_{x,+1},
    \delta_{x,-1}, 0] \times 0.1 $ & Boundary conditions,
    \eqref{eq:chemotaxis_quants}\\
    \hline
  \end{tabular}
  \begin{tablenotes}
  \item[a] Migration rate scalings are generally non-zero only to
    voxels with $u_j < u_i$, but for $D_{4}$ only migration rates to
    empty voxels with $u_j = 0$ are non-zero.
  \item[b] For $\chi_5$, $3/4$ of the cells have a uniformly random
    value between $0$ and $1$ and the rest the value $5$.
  \item[c] Kronecker delta: $\delta_{x,y} = 1$ iff $x=y$ and zero
    otherwise.
   \item[d] For models 1-4, scale $=1.044\times10^6$; model 5, $3000$.
  \end{tablenotes}
  \label{tab:parameters_chemotaxis}
  \end{threeparttable}
\end{table}

The fourth experiment uses instead the pressure-chemical coupling from
\cite{engblom2018scalable},
\begin{align}
    \label{eq:pressure-chemical_coupling}
  -\Delta p &= \nabla(p \chi \nabla s) + s(u),
\end{align}
and only pressure-driven migration. This requires a custom finite
element operator other than the Laplacian, which the solver is
capable of handling.

For models 1--3, we obtain the timescale by assuming that a cell
exposed to a chemical gradient of $0.05$, the gradient of the
underlying field without sources or sinks, moves one cell diameter per
minute, which gives the scaling of $D$, $G$, and $\xi$ above. Model 4 has
a population-dependent chemical, and, for visual comparison, we scale it
instead such that its dynamics occur over the same time interval as the
first three models. For these four models, we also assume a cell diameter
of $10\mu m$ for the spatial scale. For model 5 in 3D, we assume instead
that a unit gradient yields an expected migration rate of one cell diameter
per minute.

\begin{figure}[H]
  \centering
  \includegraphics{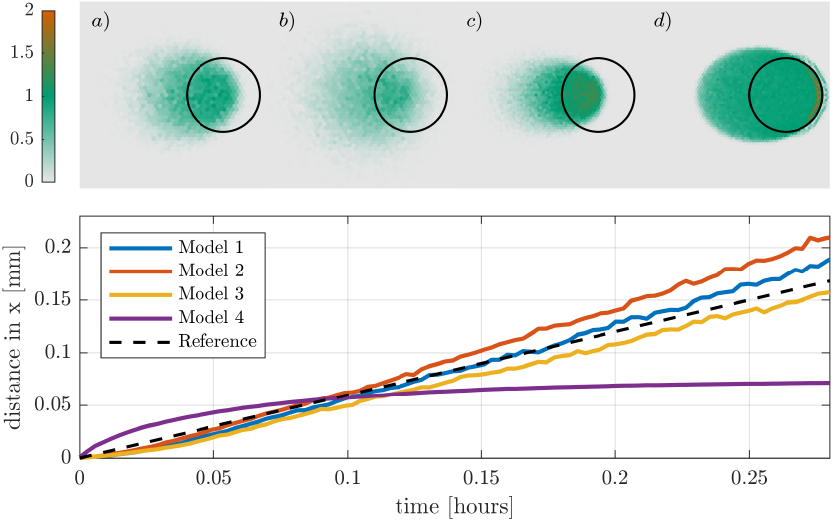}
  \caption{
    \if\submit1
    \doublespacing
    \fi
    \textbf{Four chemotaxis models in two dimensions}. Mean cell
    number over $100$ simulations of a variety of simple chemotaxis
    models. \textit{a)}: pressure-driven migration and chemotaxis;
    \textit{b)}: diffusion-driven migration and chemotaxis;
    \textit{c)}: pressure-driven migration with cells consuming a
    chemo-repellent; migration and signals defined by
    \eqref{eq:chemotaxis_rates} and
    \eqref{eq:chemotaxis_quants}. \textit{d)}: migration down the
    gradient of a field that couples the pressure and chemical signal,
    as defined by \eqref{eq:pressure-chemical_coupling}. The black
    rings denote the contour of the initial state. The bottom figure
    depicts the mean cell position over time for each model, alongside
    a reference line of a free cell moving with the expected speed of
    the chemotaxis only. The fourth model obeys different enough
    dynamics that it is not expected to follow the reference line.}
  \label{fig:chemotaxis1}
\end{figure}

\begin{figure}[H]
  \centering
  \includegraphics{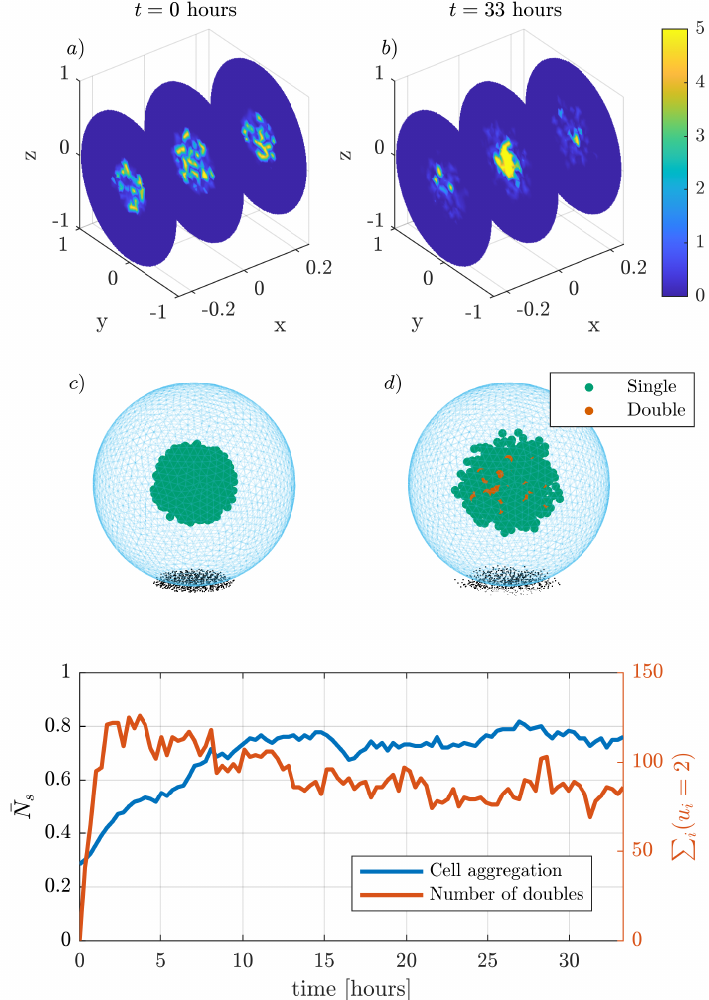}
  \caption{
    \if\submit1
    \doublespacing
    \fi
    \textbf{Chemotaxis model in three dimensions} Cells emit a
    substance that diffuses and attracts other cells proportionally to
    a cell specific signaling sensitivity $\chi$. Figures \textit{a)}
    and \textit{b)} show the sensitivity levels of cells in the
    population at $t=0$ and $t=33$ hours, respectively. Figures
    \textit{c)} and \textit{d)} show the corresponding population and
    its shadow in the $xy$ plane. Doubly occupied voxels in red. A
    small amount of cells have a significantly stronger sensitivity
    and are ultimately aggregated in the center. The bottom figure
    shows two metrics over time: aggregation of sensitive cells
    towards the center and the number of doubly occupied voxels,
    respectively.}
  \label{fig:chemotaxis2}
\end{figure}

\figref{chemotaxis1} shows the outcome of the first four models in 2D
using the corresponding four different chemotaxis models, and
\figref{chemotaxis2} shows the experiment of the final fifth model in
3D, using the parameters in \tabref{parameters_chemotaxis}. The former
2D experiments yield a notable migration in the direction of the
chemical gradient, but of clearly distinct character. The populations
of models 1--3 migrate collectively close to the speed of a single
cell moving freely by chemotaxis only. The chemotaxis in model 4,
however, is coupled to the pressure in a way that the population
approaches a steady state as the number of doubly occupied voxels goes
to zero. For the 3D experiments we measure instead $\bar{N}_s$ which
is the ratio of sensitive cells within a radius of the origin. The
radius is defined such that it spans as many voxels as there are
sensitive cells. We see that this ratio increases from $0.3$ to a
steady value $\approx 0.8$. This together with the sharp increase in
number of doubly occupied voxels signal a strong aggregation effect
due to the chemotaxis, where the less sensitive cells have been pushed
aside. This effect, alongside the cell sorting experiments in
\S\ref{sec:cellsorting}, highlight that the spatial exclusion
property, that movements into occupied voxels are generally forbidden,
does not hinder cell reorganization in the framework simulations.

These experiments showcase the flexibility of the framework and the
rich outcomes it can explore, where only a couple of changes in inputs
were required for the solver between the experiments---including the
change to 3D.

\section{Discussion}
\label{sec:discussion}


We formalized and developed the DLCM solver, an efficient
computational framework within the URDME modeling environment, for
simulating spatially stochastic dynamics in multicellular systems.  We
attained estimates of the framework's error bound compared to a
ground-truth PDE model and the framework's computational complexity
with respect to key features. These estimates and their derivation
serve as a guide for the behavior of the computational cost and error
of more complex models. Benchmark studies demonstrated the framework's
versatility and its ability to generate biologically interpretable
insights. The framework is designed to be efficient by leveraging the
use of discrete Laplacians (and related operators) for which there
exist many efficient solver alternatives; however, the current
implementation does not fully utilize this potential, instead
emphasizing here the framework's functionality and interpretability.

Designed to balance computational efficiency with detailed mechanistic
modeling, the solver supports diverse biological processes, including
chemotaxis, mechanotaxis, nutrient-driven growth, and intercellular
signaling, making it highly adaptable to various research
scenarios. Through benchmark models we stressed the capability of the
framework to investigate emergent behaviors in cell populations, in
part thanks to how it facilitates model analysis through its
continuous-physics foundation (cf.~\cite{blom2024morphological}) and
since internal states are readily modeled both discretely and
continuously. The access to a continuous-physics model counter-part is
a defining feature of the framework not immediately available to other
widely used agent-based cell population frameworks as mentioned in
\S\ref{sec:introduction}. Further, we have shown that the framework is
capable of reproducing the qualitative behavior of cell sorting
\cite{osborne2017comparing} (see \figref{cellsort}), is able to couple
a previous model for cell signaling \cite{menz2025modelling} to a
growing population (\figref{signaling_discrete},
 \ref{fig:signaling_continuous}), and can simulate a range of chemotaxis
models in both 2D and 3D (\S\ref{sec:chemotaxis}).

In summary, we regard the proposed solver as a research tool for
exploring and predicting complex multicellular dynamics, advancing the
integration of computational modeling into biological research. Its
flexibility and mechanistic foundation make it ideal for addressing
fundamental questions in cell biology.

Future efforts will focus on expanding the solver's capabilities to
model more complex cellular behaviors, enhance its computational
scalability, as well as its ability to integrate experimental data for
predictive applications.
%

The computational bottleneck of the framework is the inversion of the
discrete Laplace operator and the calculation of the curvature, if
such effects are present; this renders efficient Gillespie-variations
such as the Next Sub-volume Method \cite{elf2004spontaneous} that
optimize the event sampling process less effective.
Truly realizing the potential of the framework's efficiency requires
leveraging the highly efficient iterative solvers available for
inverting the discrete Laplace operators.
For example, using algebraic multigrid as preconditioner to iterative
methods, e.g., conjugate gradient, is a highly efficient approach for
such discrete elliptic operator problems
\cite{stuben2001amgrev}. Using optimal solvers for the Laplace
operators yields the solver computational complexity estimate in
Proposition \ref{prop:solver_complexity}.

We welcome researchers to adopt the solver also to phenomena outside
the range presented herein, e.g., in wound healing and immune
response, or for population-level behavior of more complex networks of
signaling pathways, to deepen the mechanistic understanding of
multicellular systems and drive collaborative progress in systems
biology and biomedical research. Wound healing models, for example,
might use a combination of chemotaxis and pressure- and/or
diffusion-driven migration, capabilities which were all exemplified in
\S\ref{sec:chemotaxis}. Immune response may be modeled through
additional cell types or through the micro-environment fields
depending on the application; more involved signaling pathways are
readily modeled, expanding upon or modifying the implementation used
for the experiments in \S\ref{sec:cellsignaling}.

\subsection{Availability and Reproducibility}
\label{subsec:reproducibility}

The DLCM solver is found and used with version 1.5 of the URDME
open-source simulation framework \cite{URDMEpaper}, available for
download at \url{http://www.urdme.org}. See the associated README in
the DLCM workflows folder, where the example codes are readily adapted
for other models fit for the framework.


\printbibliography[title={References}]

\appendix
\clearpage

In \S\S\ref{apx:proof}--\ref{apx:proof_estimate}, we assume some
familiarity with finite element methods (FEM), following much of the
notation from the monograph \cite{larson2013finite}, to which we refer
the reader for details. Here we use the standard function space
$V \equiv H^1(\Omegacomp) = \{v; \; \int_{\Omegacomp} \|\nabla v\|^2 <
\infty\}$ with inner product defined in \eqref{eq:inner_product} and
assume that $V_h$ is the subspace obtained using the usual piecewise
linear basis functions on the discretization $\Omega_h$ of
$\Omegacomp$, with $v = 0$ at $\partial \Omegacomp$.

\section{Proof of
  Proposition~\ref{prop:elliptic_projection_minimizer}}
\label{apx:proof}

We first show the inequality \eqref{eq:elliptic_minimizer} and then
derive the scheme \eqref{eq:elliptic_scheme}.

Let $\hat{\phi} \in V_h$ satisfy \eqref{eq:elliptic_projection} with
an arbitrary $f \in V$, and let $c \coloneqq \cell h^2$ for
brevity. Using an arbitrary $v \in V_h$ and the linearity of the inner
product, we get that
\begin{align*}
   \lVert \hat{\phi} - f \rVert^2 =  
   (\hat{\phi} - f, \hat{\phi} - v + v - f) &= (\hat{\phi} - f, \hat{\phi} -
   v) + (\hat{\phi} - f, v - f)  = \\ -c(\nabla\hat{\phi}, \nabla(\hat{\phi}
- v)) + (\hat{\phi} - f, v - f) &=
   -c(\nabla\hat{\phi}, \nabla\hat{\phi}) + c(\nabla\hat{\phi}, \nabla v)
+ (\hat{\phi} - f, v - f),
\end{align*}
using the definition of the elliptic projection
\eqref{eq:elliptic_projection}, since $\hat{\phi} - v \in V_h$. After
rearranging and using the Cauchy-Schwarz inequality twice we find that
\begin{align*}
  \lVert \hat{\phi} - f \rVert^2 +
  c \lVert\nabla\hat{\phi} \rVert^2 &=
  c(\nabla\hat{\phi}, \nabla v) + (\hat{\phi} - f, v - f) \leq \\
  c\lVert \nabla \hat{\phi} \rVert \lVert \nabla v \rVert +
  \lVert\hat{\phi} - f\rVert \lVert v - f \rVert  &\leq
                                                    \sqrt{(\lVert\hat{\phi} - f\rVert^2 +
                                                    c\lVert \nabla \hat{\phi}\rVert^2)
                                                    (\lVert v - f\rVert^2 + c\lVert \nabla
                                                    v\rVert^2)},
\end{align*}
for all $v \in V_h$. Dividing through by the square root of the left
hand side and squaring yields \eqref{eq:elliptic_minimizer}.

Finally, using the piece-wise linear basis functions
$\{\varphi_i\}_{i=1}^{n}$ over the triangulation $\Omega_h$ of the domain,
the solution to \eqref{eq:elliptic_projection} can be written as
\begin{equation*}
    \hat{\phi} = \sum_{i=1}^{n} \xi_j \varphi_j,
\end{equation*}
which, inserted back into \eqref{eq:elliptic_projection}, yields the
formulation
\begin{equation*}
    \sum_{j=1}^{n}\xi_j \left( \int_{\Omegacomp} \varphi_j \varphi_i \, dx +
    \cell h^2  \int_{\Omegacomp} \nabla \varphi_j \cdot \nabla \varphi_i \,
    dx \right) =  \int_{\Omegacomp} f \varphi_i \, dx, \quad i = 1, 2, ..., n,
\end{equation*}
which is \eqref{eq:elliptic_scheme}.

\section{Pressure Gradient Estimate}
\label{apx:proof_estimate}

We show that the finite volume (FV) scheme derived from the DLCM
migration rates \eqref{eq:rates_migration_explicit} to solve the
advection equation \eqref{eq:pressure-driven_model} is equivalent to a
first order upwind FV scheme. First, let the elements $K_i$ define a
Delaunay triangulation of $\Omegacomp$ with corresponding nodes $N_i$,
and let the dual Voronoi tesselation have volumes $V_i$. Let $u^*(t)$
be the exact solution at time $t$ to the PDE when using the perturbed
pressure $p\mapsto p_h$ defined in
\eqref{eq:pressure_perturbed_source}. To derive an underlying FV
scheme we first consider the rate of change of the average of $u^*$
over a volume $V_i$ and split the integral over the parts of the
elements $K_j$ that are in $V_i$, denoted $\tilde{K}_j$, and get that
\begin{equation}
 \begin{aligned}
      \label{eq:FV_scheme_continuous}
& \frac{1}{V_i} \int_{V_i} \frac{\partial u^*}{\partial t} =  \frac{1}{V_i}
\sum_{\tilde{K}_j \in V_i} \int_{\tilde{K}_j} \frac{\partial u^*}{\partial t} =
 \frac{1}{V_i} \sum_{\tilde{K}_j \in V_i} \int_{\tilde{K}_j} \nabla
 \cdot 
 (Du^*\nabla p_h) \\
 = \frac{1}{V_i} \sum_{\tilde{K}_j \in V_i} & \left( \int_{\partial \tilde{K}
 _j^{\mathrm{out}}} (Du^*\nabla p_h) \cdot n  + \int_{\partial \tilde{K}
 _j^{\mathrm{in}}} (Du^*\nabla p_h) \cdot n \right) = \frac{1}{V_i}  
 \int_{\partial V_i} (Du^*\nabla p_h) \cdot n, 
\end{aligned}
\end{equation}
where $\partial \tilde{K}_j^{\mathrm{out}} = \partial \tilde{K}_j \bigcap
\partial V_i$ such that $\partial \tilde{K}_j^{\mathrm{in}}$ are edges internal
to $V_i$. Flow over boundaries inside $V_i$ does not change the
volume-average of $u^*$, which gives the final equality in
\eqref{eq:FV_scheme_continuous}. 

From \eqref{eq:FV_scheme_continuous} we can construct a
\emph{donor-cell upwind} FV scheme
\cite[Sect.~20.1]{leveque2002finite} for $u^*$ but with a discrete
estimate of the velocity field using the nodal pressure values as
follows. Let $u_h(t)$ be the semi-discrete FV solution at timestep $t$
with $u_h(N_i) \coloneqq u_i = V_i^{-1}\int_{V_i}u_h dV + \Ordo(h^2)$
satisfying
      \begin{equation}
      \label{eq:DLCM_FV_scheme}
 \frac{\partial u_i}{\partial t} = \frac{1}{V_i}
\sum_jD\frac{p_j - p_i}{h_{ij}}e_{ij} R_{ij} + \Ordo(h^2),
\end{equation}
where $p_i = p_h(N_i)$ for node $N_i$, $V_i$ the voxel volume, and
$R_{ij}$ an upwind indicator function equal to $u_i$ when $p_j-p_i$ is
negative and $u_j$ else in an upwind fashion. Note the similarity
between \eqref{eq:DLCM_FV_scheme} and
\eqref{eq:rates_migration_explicit},

Consider the pressure gradient along any edge,
$\nabla p_h \cdot n \vert_{e_{ij}}$. Thanks to the linearity of $p_h$
we can derive an exact expression of its gradient at the edge
$e_{ij}$, using the decomposition $p_h = \sum_k p_k \phi_k$ where
$\phi_k$ is the linear basis function corresponding to node $N_k$. We
assume that any edge $e_{ij}$ is contained within the two elements
shared by the nodes $N_i$ and $N_j$. The Voronoi property implies that
$e_{ij}$ is perpendicular to $N_j-N_i$ (the line segment between the
nodes in question), and together with the assumption on edges we get
that $\nabla \phi_k \cdot n = \pm h_{ij}^{-1}$ across the entire edge
for $k = i, j$, and zero for all other basis functions. Therefore,
\begin{equation}
\label{eq:pressure_gradient_approximation}
 \nabla p_h \cdot n\vert_{e_{ij}} =   \sum_k p_k \nabla
 \phi_k \cdot n \vert_{e_{ij}} = \frac{p_j - p_i}{h_{ij}}.
\end{equation}
In other words, the pressure gradient over each edge is constant and
exactly equal to the estimate in \eqref{eq:DLCM_FV_scheme} and the
standard results for upwind finite element schemes apply making it
order one in space (cf.~\cite[Sect.~20.1]{leveque2002finite} for
Cartesian grids). It should be remarked that the accuracy of FV
schemes over unstructured meshes is non-trivial to analyze,
particularly near boundaries \cite{FVMaccuracy}.

\section{Proof of Proposition~\ref{prop:solver_complexity}}
\label{apx:proof_complexity}

We derive the DLCM solver computational complexity under assumed
bounds on the event intensities and the number of cells. We start by
deriving the complexity of using an optimal Laplace solver after every
event in the external layer, and later add the complexity of the
independent internal dynamics.

Consider the time interval $(0,T)$ where the number of cells is
bounded by $\Ncells$ and let $\mathcal{W}$ and $\mathcal{V}$ be the
bounds $\sum_r \omega_r(u_i,v_i,w_i) \leq \mathcal{W}$ and
$\sum_r \nu_r(u_i,v_i,w_i) \leq \mathcal{V}$ for all cells $i$ during
this interval. The total intensity of the external events is then
$\le \mathcal{I} + \mathcal{W}\Ncells$ where $\mathcal{I}$ is the
total pressure-driven migration intensity for the population. This
bound also serve as an upper bound for the mean of $1/dt$. The
expected number of events during $(0,T)$ is $T/dt$ and, using an
optimal Laplace solver over the grid of voxels with
$\Ordo(\Nvox)$ operations per external event, the total expected
complexity becomes bounded by
$T/dt \times\Ordo(\Nvox) = \Ordo (T \times [\mathcal{I} +
\mathcal{W}\Ncells]\Nvox)$.

We estimate $\mathcal{I}$ starting with the total flux in or out of each 
voxel,
\begin{equation}
\left| \int_{\partial \Omega_i}  \nabla p \cdot n \ d\Omega \right| = 
\left| \int_{\Omega_i}  \Delta p \ d\Omega \right| = \lvert s^{(0)}(u_i, 
v_i, w_i)\rvert \lvert \Omega_i \rvert.
\end{equation} 
With $s^{(0)}$ uniformly bounded, we sum the intensities over all
voxels and get that the intensity $\mathcal{I}$ -- which excludes
surface tension effects -- is $\Ordo (\Ncells\Nvox^{-1})$,
understanding that $\lvert \Omega_i \rvert \propto \Nvox^{-1}$, i.e.,
that the volume is bounded. A similar analysis can be performed on
more complicated migration potentials involving diffusive quantities
defined by \eqref{eq:laplace_quantities}.

Surface tension, on the other hand, induces at most a pressure
gradient proportional to the smallest possible curvature (half a cell
radius) at each boundary node and towards each neighboring node
containing the same cell type: the added intensity is proportional to
$h^{-1} \propto \Nvox^{1/d}$, where $d$ is the number of spatial
dimensions. Since the total number of boundary nodes is bounded by
$\Ncells$ -- clearly a worst case estimate -- we get that the surface
tension contributes to the complexity by $\Ordo\left(\sigma_{\max}
\Nvox^{1/d}\Ncells\right)$, where $\sigma_{\max} \equiv \max_{kl}\sigma_{kl}$.

Finally, each external event also entails independent calculations of
the internal event dynamics, with mean intensity
$1/d\tau \le \mathcal{V}$ per cell.  Thus, with the expected number of
such events during the full interval being $T/d\tau$, the expected
complexity for the internal events becomes
$T/d\tau \times\Ordo(\Ncells) = \Ordo (T \mathcal{V}
\Ncells)$. Since the internal and external events are independently
executed, we sum their respective expected complexity to get the full
solver complexity bound \eqref{eq:solver_complexity}.

\end{document}